\documentclass[11pt]{article}
\usepackage{amsmath}
\usepackage{graphicx}
\usepackage{bm}
\usepackage{xcolor}
\usepackage{amssymb}
\usepackage{amsfonts}
\usepackage{comment}
\usepackage{cite}
\usepackage{todonotes}
\usepackage{caption}
\usepackage{subcaption}
\usepackage{CJKutf8}
\usepackage{authblk}

\usepackage[utf8]{inputenc}

\def\be{\begin{equation}}
\def\ee{\end{equation}}
\def\ba{\begin{eqnarray}}
\def\ea{\end{eqnarray}}
\def\nn{\nonumber}

\def\bl#1\el{\begin{align}#1\end{align}}

\title{Gravitational effects on Hong-Ou-Mandel interference in terrestrial laboratory }

\author[a,b]{Xuan Ye\thanks{ yexuan@cuhk.edu.cn}}
\author[b]{Yang Zhang \thanks{ yzh@ustc.edu.cn}} 
\author[b]{Bo Wang \thanks{ ymwangbo@ustc.edu.cn}} 
\affil[a]{\small School of Science and Engineering, The Chinese University of Hong Kong, 
Shenzhen, 518172, Guangdong,China}
\affil[b]{\small Department of  Astronomy,  Key Laboratory
for Researches in Galaxies and Cosmology,
 University of Science and Technology of China,  Hefei, Anhui, 230026,  China}

 \date{}

 \date{}

\def\be{\begin{equation}}
\def\ee{\end{equation}}
\def\ba{\begin{eqnarray}}
\def\ea{\end{eqnarray}}
\def\nn{\nonumber}
\def\bl#1\el{\begin{align}#1\end{align}}

\large

\begin{document}
\begin{CJK}{UTF8}{gbsn}
\maketitle
\allowdisplaybreaks
\begin{abstract}
In this study, we investigate how Earth's gravitational field affects  Hong-Ou-Mandel (HOM) interference 
experiments conducted in a terrestrial laboratory.
To second order, we calculate the relativistic time delay from the null geodesic equation (particle perspective), while the phase shift and the associated effective time delay are derived from the Klein-Gordon equation (wave perspective). Since gravity influences both the temporal and spatial parts of the phase shift, these two time delays differ and lead to different coincidence probabilities. The previous HOM 
experiment conducted on a 
rotating platform suggests that the wave perspective can explain the experimental results. 
We further explore the frame dragging and redshift effects in an arbitrarily oriented rectangular interferometer under two distinct scenarios with different photon paths, measuring one effect in each scenario. We find that both effects can be amplified by increasing the number of light loops. Additionally, we emphasize that the next-to-leading order Sagnac effect, arising from gravitational acceleration, is comparable to the Thomas precession, the geodetic effect, and the Lense-Thirring effect. To detect the leading order Sagnac effect and the redshift effect caused by gravitational 
acceleration, we estimate the number of loops that photons should travel in the 
interferometer. Furthermore, we propose that the difference between two HOM patterns can be used as a probe to detect gravitational effects on quantum systems.
\end{abstract}

\section{Introduction}

In recent decades, fueled by the continuous advancement of quantum technologies \cite{Milburn2003,peres200d4,yin2017satellite,liao2017satellite,ren2017ground,xu2019satellite,li2022quantum,Bowangm}, there has been growing interest in observing gravitational effects in quantum experiments. These experiments help us understand the behavior of quantum systems in gravitational fields. Refs.\cite{Pound1,Pound2} measured the gravitational redshift of 
$\gamma$-rays using nuclear resonance.
Refs. \cite{Brushi20141, Brushi20142,Brischi3,Brischi4,Brischi5} explored the gravitational effects on wave packets of coherent light and their  applications in quantum communication. The Colella, Overhauser, and Werner (COW) experiment employed neutron interference to test the non-relativistic gravitational effects on matter wave functions \cite{Colella1975}. Furthermore, the optical COW experiments proposed in Refs.\cite{Hilweg, Zych1, Zych2} aim to detect the phase shift caused by Earth's gravitational field in the wave function of a single photon. In addition to single-photon interference experiments, the two-photon interference experiment known as Hong-Ou-Mandel (HOM) interference is also a  potential candidate for studying the influence of gravity on quantum systems.

The HOM interference experiments involve the indistinguishability of two photons without a classical analogue \cite{Branczyk2017, Ou2007}. An intriguing phenomenon in HOM interference experiments is photon bunching: when two indistinguishable photons simultaneously enter a 50:50 beam splitter, quantum interference causes the photons to bunch together and exit through the same output port, resulting in the suppression of coincident detections at both output ports. This phenomenon reflects the bosonic property of photons, as observed in the pioneering experiment \cite{Hong1978}. The HOM interference, serving as a natural tool to measure gravitational effects in quantum systems, has been studied from various perspectives. Refs.\cite{Brazel2,Brushi20141, Brushi20142} predicted the difference in the HOM interference pattern between a terrestrial laboratory and a satellite based on three spectral profiles, assuming that Earth is a non-rotating symmetric object. Ref.\cite{Brandy2021} considered the HOM experiment within a terrestrial laboratory and studied the relativistic influences caused by frame dragging and redshift effects on the HOM pattern. Ref.\cite{Kish2022} used a rotating turntable as an analogy to the Kerr black hole to discuss the frame dragging effect in HOM interference experiments. On the experimental side, Ref.\cite{Restuccia2019} was the first to observe the influence of the Sagnac effect in the HOM interference experiment using a rotating platform. Refs.\cite{Silvestri20221, Silvestri20222, Silvestri2023} used a 2 km fiber to build a Sagnac interferometer with an effective area of approximately 750 m$^2$ and detected the Sagnac effect caused by the rotation of Earth. Ref. \cite{Mohageg2022} suggested exploring HOM interference experiments in deep space. In Ref. \cite{HuiNan}, the authors examined the key technologies over an 8.4 km free-space channel to conduct the HOM experiment in a satellite-based system.

The key quantity in the HOM interference experiments is the coincidence probability, which represents the interference pattern as a function of the time delay. In the literature, two perspectives are often used to encode gravitational effects into the HOM interference pattern: the particle perspective and the wave perspective. In the particle perspective \cite{Brushi20141, Brushi20142, Restuccia2019, Kish2022}, the coincidence probability is calculated using the two-photon state constructed from the relativistic time delays obtained by solving the null geodesic equation. However, from the wave perspective, the coincidence probability is calculated using the two-photon state constructed from the effective time delays, which are derived from the phase shift obtained by solving the Klein-Gordon equation \cite{Brandy2021,Caiyiqi, Leo, Canad, Jerzy}. We calculate the time delays based on these two perspectives up to the second order and find that they differ by several terms at the second order and even by a minus sign at the first order. These differences arise because gravity affects both the temporal and spatial parts of the phase shift. Nevertheless, in the context of the HOM experiments, the time delays from these two perspectives have not been carefully distinguished. The analysis of the previous HOM experiment conducted on a rotating platform \cite{Restuccia2019} suggests that the experimental results can be explained from the wave perspective. This suggests that when considering a photon system in the framework of general relativity, the wave picture of the photon is more appropriate than the particle picture.
 
In this paper, we study the redshift and frame dragging effects caused by Earth's gravitational field on the HOM interference patterns in a terrestrial laboratory from the wave perspective. 
We discuss the feasibility of 
performing canonical quantization on a massless minimally-coupling scalar field in a terrestrial laboratory 
and show the connection between operators, mode functions, and quantum states, which has 
not been extensively covered in the literature. Moreover, we allow the areal vector and arm vectors 
of the interferometer to point in arbitrary directions, thereby generalizing the work of 
Ref.\cite{Brandy2021}. We examine the time delays within the same interferometer configuration in two specific scenarios involving different light paths. In the first scenario, the redshift effect is absent, 
and only the frame dragging effect remains. Conversely, in the second scenario, the frame dragging effect 
is absent, leaving only the redshift effect. This approach enables us to study the independent 
influence of these two effects on the coincidence probability. Additionally, we estimate the number
of loops photons should travel in the interferometer for each scenario to ensure that 
the Sagnac effect and the gravitational acceleration-induced redshift effect are comparable to the current
experimental accuracy of
$\sim 10^{-18}$s, while keeping the interferometer size remains manageable. 
We also introduce a probe, the difference between two coincidence probabilities with 
different effective time delays, to detect the relativistic influences and compare it with that from the other interferometer described in Ref.\cite{Brazel2}.

This paper is organized as follows. In Sec. \ref{ref1}, we review the formulas for the HOM interference experiments in {Minkowski} spacetime and the relationships between quantities in the laboratory and the geocentric reference frame. In Sec. \ref{homdae}, we study the coincidence probability encoding the redshift and frame dragging effects from two different perspectives. In Sec. \ref{expsetup}, we present the interferometer configuration. In Sec. \ref{analysetimedelay}, we study the time delays caused by the frame dragging and redshift effects in detail. In Sec. \ref{interference}, we revisit the  previous HOM experiment performed on a rotating platform and examine how the gravitational field of Earth influences the HOM interference patterns with various spectral profiles. Sec. \ref{conclusionanddis} gives the conclusions and {discussion}. Appendix \ref{Appanlase} contains the values used in this paper. Appendix \ref{Appcoordinate} shows the vectors spanned by various basis vectors. Appendix \ref{APPad} exhibits the process of connecting the quantities in the geocentric reference frame to those of the laboratory reference frame. Appendix \ref{Appcontoti} connects the relativistic time delay to the phase shift up to the second order. Appendix \ref{teimdelayvar} outlines the computation of the time delays resulting from various gravitational effects.

\section{Basics}\label{ref1}

To clarify the {notation}, in this section we briefly review the theoretical framework for predicting the statistics of the HOM experiments in Minkowski spacetime \cite{Brazel1, Brazel2}, and the relations between the quantities in the laboratory reference frame and those in the geocentric reference frame \cite{Delva2017}.

\subsection{Hong-Ou-Mandel interference}\label{HOMinter}

\begin{figure}[htb]
 \centering
 \subcaptionbox{}
   {%
     \includegraphics[width = .35\linewidth]{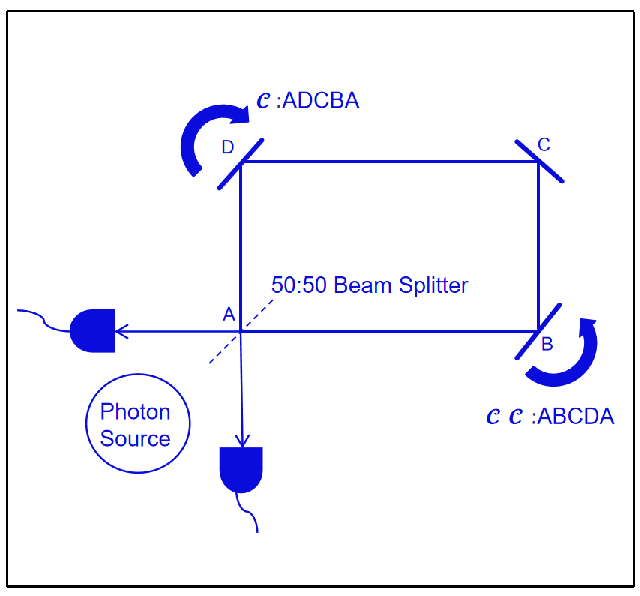}}
      \hspace{.50in}
      \subcaptionbox{}
   {%
     \includegraphics[width = .342
\linewidth]{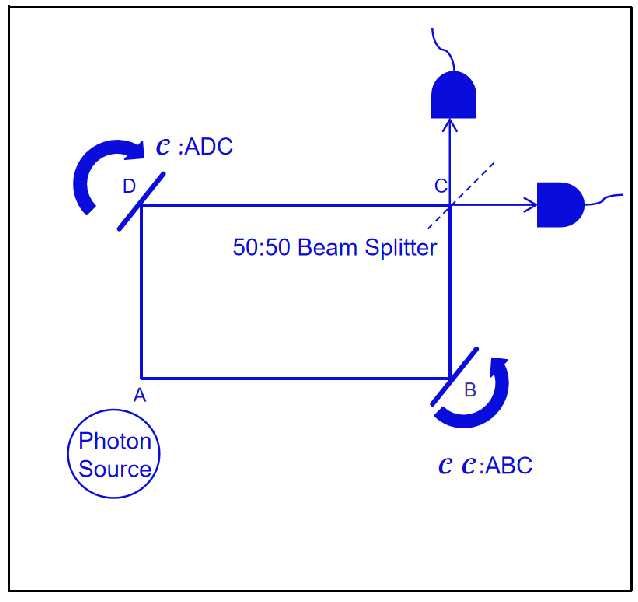}}  
\caption{
Schematic diagrams of interferometric optical paths. (a) Both the photon source and the 50:50 beam splitter (dashed line) are located at $A$. One photon travels through the counterclockwise path, ${\cal CC:} ABCDA$, while the other travels through the clockwise path, ${\cal CC:} ADCBA$. These paths are used to detect the frame dragging effect. (b) The photon source is  located at $A$ and the 50:50 beam splitter (dashed line) is located at $C$. One photon follows the counterclockwise path, ${\cal CC:} ABC$, and the other follows the clockwise path, ${\cal CC:} ADC$.  These paths are used to detect the redshift effect.} \label{intermoreferconfing}
\end{figure}

In the HOM interference experiment, two photons are emitted from the same photon source and reach the 50:50 beam splitter (BS) at different times, traveling along separate paths. Subsequently, they interfere at the BS before reaching the detectors. Schematic diagrams illustrating the interference configurations are shown in Figs. \ref{intermoreferconfing}(a) and (b). The primary physical quantity measured in the experiment is the probability that each detector detects one photon, called the coincidence probability. We shall derive this quantity in what follows.

In Minkowski spacetime, a general two-photon state in the context of the HOM interference
experiment is given by \cite{Brazel1,Pradana2019,Ou2006,Grice1997}
\bl
\left|\phi\left(t_1, t_2\right)\right\rangle=
\sum_{\boldsymbol{\sigma}_1,\boldsymbol{\sigma}_2}\mathcal{N}_\phi \int d \omega_1 d \omega_2 
\Phi\left(\omega_1, \omega_2,\boldsymbol{\sigma}_1,
\boldsymbol{\sigma}_2\right) e^{i \omega_1 t_1} 
e^{i \omega_2 t_2} a_{\boldsymbol{\sigma}_1}^{\dagger}(\omega_1) 
a_{\boldsymbol{\sigma}_2}^{\dagger}(\omega_2)|0\rangle,\label{generaleq1}
\el
{where $\omega_1$ and $\omega_2$ are the central 
frequencies of two photons, and $\boldsymbol{\sigma}_1$ and $\boldsymbol{\sigma}_2$ represent
their respective degrees of freedom}, such as polarization, path, etc. 
$\Phi$ is the joint spectral amplitude (JSA)
function describing the two-photon 
system. $t_1$ and $t_2$ are the travel times of the two photons from emission to interference.
${\cal N}_{\phi}$ is the normalization constant ensuring 
$\langle\phi\left(t_1, t_2\right)\left|\phi\left(t_1, t_2\right)\right\rangle=1$. 
$a_{\boldsymbol{\sigma}}^{\dag}(\omega)$ is
the creation operator ($\boldsymbol{\sigma}=(\boldsymbol{\sigma}_1,\boldsymbol{\sigma}_2)$,
$\omega=(\omega_1,\omega_2)$). $|0\rangle$  is the vacuum state defined by 
$a_{\boldsymbol{\sigma}}(\omega)|0\rangle=0$, where $a_{\boldsymbol{\sigma}}(\omega)$ 
is the annihilation operator. The creation and annihilation operators
satisfy the commutation relation
\bl
[a_{\boldsymbol {\sigma}}(\omega),
a_{\boldsymbol{ \sigma'}}^{\dag}(\omega')]
=\delta_{\boldsymbol{\sigma},\boldsymbol{\sigma'}}\delta(\omega-\omega'), \label{commuation}
\el
where $\delta_{\boldsymbol{\sigma},\boldsymbol{\sigma'}}$ 
denotes the Kronecker delta $\delta_{\boldsymbol{\sigma}\boldsymbol{\sigma'}}$
for discrete $\boldsymbol{\sigma}$ and the Dirac delta function 
$\delta(\boldsymbol{\sigma}-\boldsymbol{\sigma'})$  for continuous $\boldsymbol{\sigma}$.

Similar to Glauber’s theory of the
joint detection of two electric fields
\cite{Grice1997,Glauber}, the probability of the joint detection of two photons is given by
\bl
\langle\phi\left(t_1, t_2\right)|
a_{\boldsymbol{\sigma}_1}^{\dag}(\tau_1)
a^{\dag}_{\boldsymbol{\sigma}_2}(\tau_2)
a_{\boldsymbol{\sigma}_2}(\tau_2)
a_{\boldsymbol{\sigma}_1}(\tau_1)
|\phi\left(t_1, t_2\right)\rangle,
\label{jointdetection}
\el
which represents the probability that two detectors detect photons 
with $\boldsymbol{\sigma}_1$ and $\boldsymbol{\sigma}_2$ at times $\tau_1$ and $\tau_2$. 
The creation operator $a^{\dag}_{\boldsymbol{\sigma}}(\tau)$ relevant to the detector is  
defined as 
\bl
a_{\boldsymbol{\sigma}}^{\dag}(\tau)&=
\int d\omega e^{i\omega \tau} a^{\dag}_{\boldsymbol{\sigma}}(\omega), \label{eqution3}
\el 
the annihilation operator $a_{\boldsymbol{\sigma}}(\tau)$ can 
be obtained by taking the Hermitian conjugate of \eqref{eqution3}. 
Substituting \eqref{generaleq1} and \eqref{eqution3} 
into \eqref{jointdetection}, using 
\eqref{commuation}, and integrating over
$\tau_1, \tau_2$ in \eqref{jointdetection}, one gets
the unnormalized coincidence probability \cite{Brazel1,Brazel2}
\bl
P_{\boldsymbol{\sigma}_1 ,\textbf{}\boldsymbol{\sigma}_2}\left(t_1, t_2\right)
  &=2\pi|\mathcal{N}_\phi|^2\Big\{
  \int d \omega_1 d \omega_2
  \Big[ |\Phi\left(\omega_1, \omega_2,\boldsymbol{\sigma}_1,
  \boldsymbol{\sigma}_2\right)|^2
  +|\Phi\left(\omega_2, \omega_1,\boldsymbol{\sigma}_2,
  \boldsymbol{\sigma}_1\right)|^2\Big]\nn
  \\
  &~~~~~~~+2\text{ Re}\big[\int d \omega_1 d \omega_2\Phi^*
  \left(\omega_2, \omega_1,
  \boldsymbol{\sigma}_2,
  \boldsymbol{\sigma}_1\right)\Phi\left(\omega_1, \omega_2,\boldsymbol{\sigma}_1,
  \boldsymbol{\sigma}_2\right)
  e^{i (\omega_1-\omega_2) (t_1-t_2)} \big]\Big\},\label{jointprperwrittingpaper}
\el 
which differs from that in Refs.\cite{Brazel1,Brazel2} by a factor of $2\pi$ due to 
our use of the identity $\int e^{itx} dt=2\pi\delta(x)$.
 The coincidence probability \eqref{jointprperwrittingpaper} 
 depends on $(t_1-t_2)$, the difference of
  the travel times of the two photons. When gravity is not 
  taken into account, 
$(t_1-t_2)$ can be controlled  by changing the optical path difference between 
the two photons \cite{Branczyk2017,Restuccia2019}. However, 
when the length scale of the HOM interference experiment is large,
gravity will contribute an additional time delay 
in $(t_1-t_2)$ and further affect the coincidence probability. 
This effect is the main focus of this study and will be analyzed in the following sections.

In what follows, we calculate the coincidence probability by determining the quantum state of the two photons after passing through the beam splitter (BS) and the corresponding JSA. The two photons follow different light paths in Figs. \ref{intermoreferconfing} (a) and (b), which are used to detect distinct gravitational effects discussed in Sec. \ref{analysetimedelay}. The abstract formalism for computing the coincidence probability is independent of the specific light paths. Therefore, before Sec. \ref{analysetimedelay}, we simply use ${\cal CC}$ and ${\cal C}$ to denote the counterclockwise and clockwise paths, respectively, without specifying the particular light paths they represent.

Before passing through the BS, the paths of the photons are deterministic, and the corresponding quantum state is
\bl
\left|\phi(t_1,t_2)\right\rangle=
  & \int d \omega_1 d \omega_2 \phi\left(\omega_1, \omega_2\right)
   e^{i \omega_1 t_1} e^{i \omega_2 t_2}
   a_{\cal CC}^{\dagger} (\omega_1)a_{\cal C }^{\dagger}(\omega_2)|0\rangle,\label{Phione}
\el
which is obtained by using $\boldsymbol{\sigma}_1 = {\cal CC}$, $\boldsymbol{\sigma}_2 = {\cal C}$, and $\Phi\left(\omega_1, \omega_2, {\cal CC}, {\cal C} \right) = \phi(\omega_1, \omega_2)$ in \eqref{generaleq1}, where $\phi(\omega_1, \omega_2)$ is the spectral profile determined by the photon source. After passing through the BS, each photon has a 50\% probability of continuing along its original path and a 50\% probability of diverting onto another path. The paths of the two photons are no longer deterministic. The two-photon state after passing through the BS can be obtained by replacing creation and annihilation operators in \eqref{Phione} as follows \cite{Branczyk2017,Bachor2004}
 \begin{subequations} 
\bl
\hat{a}^{\dag}_{\cal C}(\omega)
 \rightarrow e^{i\eta}\hat{a}_{\cal C}^{\dag}(\omega)-\hat{a}_{\cal CC}^{\dag}(\omega),
\\
\hat{a}^{\dag}_{\cal CC}(\omega)
 \rightarrow \hat{a}_{\cal C}^{\dag}(\omega)+e^{-i\eta}\hat{a}_{\cal CC}^{\dag}(\omega),
 \el
 \end{subequations}
where the phase factor $\eta$  depends on the specific material of 
the BS and  does not affect the coincidence probability.
Thus, the two-photon state after passing through the BS can be written as 
\bl
\left|\phi(t_1,t_2)\right\rangle
  & =\int d \omega_1 d \omega_2 
   e^{i \omega_1 t_1} e^{i \omega_2 t_2} 
   \phi\left(\omega_1, \omega_2\right)\left(\begin{array}{cc}
    \hat{a}_{\cal C}^{\dagger}(\omega_1) 
    & \hat{a}_{\cal CC}^{\dagger} (\omega_1)
    \end{array}\right)\left(\begin{array}{cc}
      e^{i\eta} & 1 \\
      -1 & -e^{i\eta}
      \end{array}\right)\left(\begin{array}{cc}
      \hat{a}_{\cal C }^{\dagger}(\omega_2) \\
      \hat{a}_{\cal CC }^{\dagger}(\omega_2)
  \end{array}\right)|0\rangle.\label{psif}
\el
The JSA function in \eqref{jointprperwrittingpaper} can be directly read off 
from \eqref{psif} as
\bl
\Phi\left(\omega_1, \omega_2,\boldsymbol{\sigma}_1,
\boldsymbol{\sigma}_2\right)=\phi(\omega_1,\omega_2)
\left(\begin{array}{cc}
  e^{i\eta} & 1 \\
  -1 & -e^{i\eta}
  \end{array}\right), \label{spectrum}
\el
where both $\boldsymbol{\sigma}_1$ and $\boldsymbol{\sigma}_2$ belong to the set $\{\cal CC,\cal C\}$.

The normalized coincidence probability that each detector detects one particle is represented as  
\bl
P^{c}(t_1 - t_2) &\equiv \frac{P_{\cal C, CC} + P_{\cal CC, C}}{P_{\cal C, C} + P_{\cal C, CC} + P_{\cal CC, C} + P_{\cal CC, CC}} \nn \\
&= \frac{1}{2} \Bigg\{ 1 - \frac{ 
  2 \operatorname{Re} \Big[ \int d \omega_1 d \omega_2 \, 
  \phi^*(\omega_2, \omega_1) \,
  \phi(\omega_1, \omega_2) \,
  e^{i (\omega_1 - \omega_2)(t_1 - t_2)} \Big] }{
  \int d \omega_1 d \omega_2 \left( 
    |\phi(\omega_1, \omega_2)|^2 + |\phi(\omega_2, \omega_1)|^2 
  \right)
} \Bigg\}, \label{eq27}
\el
where  \eqref{jointprperwrittingpaper} and \eqref{spectrum} have been used. 
The coincidence probability in \eqref{eq27} is a function of $(t_1 - t_2)$. 
The normalization constant ${\cal N}_{\phi}$ is irrelevant to the HOM interference experiments because it cancels out in \eqref{eq27}. Therefore, it can be initially omitted in \eqref{generaleq1}.

We study two spectral profiles in this paper.  
The first is the original HOM profile as described in Ref.~\cite{Hong1978}, and the second is the quantum beating profile as described in Ref.~\cite{Hong1988}.  In the original HOM case, the photons are spectrally indistinguishable and frequency uncorrelated,
the spectral profile is \cite{Brazel1,Brazel2}
\bl
\phi(\omega_1,\omega_2)
=\frac{1}{\sqrt{2\pi}\zeta}e^{-\frac{(\omega_1-\nu)^2+(\omega_2-\nu)^2}
{4\zeta^2}},\label{spectrumHOM}
\el
where $\nu$ is the deviation from the center frequency and 
$\zeta$ is the single-photon bandwidth. 
Substituting \eqref{spectrumHOM} into \eqref{eq27} yields
\bl
P^c_{\text{HOM}}(t_1-t_2)=\frac12[1-e^{-\zeta^2(t_1-t_2)^2}],\label{jointcoindeiProHOM}
\el
which is the well-known formula for the HOM coincidence probability.

In the quantum beating case, the photons are spectrally indistinguishable and frequency-entangled, and the spectral profile is \cite{Brazel1,Brazel2}
\bl
\phi_{\text {QB }}\left(\omega_1, \omega_2\right)=\mathcal{N}\left[\phi_{\text{f.d}}(\omega_1, \omega_2)+e^{i \varphi} \phi_{\text{f.d}}(\omega_2, \omega_1)\right],
\label{quspectrum}
\el
where the subscript QB denotes the quantum beating, and 
\bl
\phi_{\text{f.d}}\left(\omega_1, \omega_2\right)&=\sqrt{\frac{2}{\pi \zeta}} 
\delta\left(\omega_p-\omega_1-\omega_2\right) 
\operatorname{sinc}\left(\frac{\omega_1-\omega_2-\mu}{\zeta}\right),\label{phifromfd}
\el
where $\omega_p$ is the pump frequency of the down-conversion process, and the subscript f.d. denotes frequency detuning.  
The parameter $\mu$ is the separation between the center frequencies of the two photons, and $\text{sinc}(x) \equiv \sin(x)/x$.

 Substituting
\eqref{phifromfd} into \eqref{eq27} yields  
\bl
P^c_{\text{QB}}(t_1-t_2)=\frac12\Big(1-2 \mu \mathcal{N}^2[R_{\mu \zeta}^{\varphi}\left(t_1, t_2\right)+S_{\mu \zeta}\left(t_1, t_2\right)]\Big),\label{jointcoindeiProQB}
\el
where 
\begin{subequations} 
\bl
{\cal N}^{2}&=\Big(2 \mu(1+\cos\varphi \operatorname{sinc}(2 \mu / \zeta))\Big)^{-1},
\\
R_{\mu\zeta}^{\varphi}\left(t_1, t_2\right)&=\cos (\mu (t_1-t_2)-\varphi) \operatorname{tri}\Big(\frac{\zeta (t_1-t_2)}{2}\Big),
\\
 S_{\mu\zeta}\left(t_1, t_2\right)&=\sin \Big(\frac{2 \mu}{\zeta} \operatorname{tri}\big(\frac{\zeta (t_1-t_2)}{2}\big)\Big)/\frac{2 \mu}{\zeta},
\el
and 
\bl
\operatorname{tri}(x)= \begin{cases}1-|x| & \text { if }|x| \leq 1 
  \\ 0 & \text { if }|x|>1\end{cases},
\el
is the triangular function.
\end{subequations} 

The coincidence probabilities \eqref{jointcoindeiProHOM} and \eqref{jointcoindeiProQB} are valid in Minkowski spacetime. To show how the gravitational field of Earth affects these coincidence probabilities, we  need to determine the relations between the metric of the laboratory reference frame and that of the geocentric reference frame. This will be discussed in the following subsection.

\subsection{Relate the laboratory reference frame to the geocentric reference frame}\label{sect3}

We consider the HOM experiments conducted in a terrestrial laboratory rather than satellite-based experiments \cite{Zych1,Zych2,Brushi20141,Brushi20142,Terno2020,Terno2021}.  
The proper reference frame (i.e., the laboratory reference frame) \cite{MTW} is a natural choice for studying the HOM experiments under Earth's gravitational field \cite{booksoffel1,booksoffel2,Ciufolini}. However, it is more convenient to analyze Earth's gravitational field in the geocentric reference frame. In this subsection, we shall provide the relations between quantities in the laboratory reference frame and those in the geocentric reference frame.

The metric along the worldline of an accelerated observer in the laboratory reference frame takes the form
$g_{\mu\nu}=\eta_{\mu\nu}+h_{\mu\nu}$, where $\eta_{\mu\nu}$ is the Minkowski metric and $h_{\mu\nu}$  is the metric perturbation \cite{MTW,booksoffel1,booksoffel2,Ciufolini,Ni}. The corresponding 
line element $ds$ in Cartesian coordinates $x^{\mu} = (x^0, x^i)$  is given by 
\bl
d s^2&=g_{\mu\nu}dx^{\mu}dx^{\nu}
=-[1+2\frac{\vec{\gamma}\cdot \vec{x}}{c^2} 
+\frac{1}{c^4} (\vec{\gamma}\cdot \vec{x})^2
+\frac{1}{c^2} (\vec{\omega}'\cdot \vec{x})^2
-\frac{1}{c^2}\vec{\omega}'^2 \vec{x}^2
+R_{0l0m} x^{l}x^m] (d x^{0})^2\nn
\\
&~~~~~~~~~~~~~~~~~~+[\frac{\vec{\omega}^{\prime} \times \vec{x}}c- 
\frac{2}{3} R_{0lim}x^lx^m] (d\vec{x}) (d x^{0})+[\delta_{ij}-R_{iljm}x^lx^m]
d x^i dx^j,\label{observrefraenfram12}
\el
where $c$ is the speed of light.  The effect associated with $h_{00}$ is known as the ``redshift effect'', and that associated with $h_{0i}$ is known as the ``frame dragging effect'' \cite{Brandy2021}. The vectors $\vec{\gamma}$ and $\vec{\omega}'$ denote the acceleration and angular velocity of the laboratory, respectively. Both are constants and can be measured by accelerometers and gyroscopes \cite{MTW,Brandy2021}. $R_{\mu\nu\alpha\beta}$ is the Riemann tensor evaluated at the origin of the laboratory reference frame. It can be obtained by contracting the tetrad $e^{\mu}_{~(\alpha)}$, defined in Appendix \ref{APPad}, with the Riemann tensor computed in the geocentric reference frame. For further details, see Ref.~\cite{will222}. The explicit expression of the Riemann tensor is not listed here, as it will not be used. To calculate the second order time delays in the next section, the metric in \eqref{observrefraenfram12} is expanded up to second order in $x^{\mu}$.

The vectors $\vec{\gamma}$ and $\vec{\omega}'$ {can be expressed as functions of} the gravitational field of Earth, 
which is usually
described by the post-Newtonian metric  
in the geocentric reference frame\cite{Cwill,Erick,booksoffel1,booksoffel2,Ciufolini},
\bl
ds^2&=\mathfrak{g}_{\mu\nu}dX^\mu dX^\nu=-(1-\frac{2U}{c^2}+\frac{2U^2}{c^4})(dX^0)^2
-8\frac{V_i}{c^3}(dX^i)(dX^0)+(1+\frac{2U}{c^2})\delta_{jk}(dX^j)(dX^k), \label{PNmetric}
\el
where $X^{\mu}=(X^0,X^i)$ are the Cartesian coordinates. The 
gravitational potential
$U$ is defined as 
\bl
U(|\vec{X}|)=\frac{G M}{|\vec{X}|},\label{gravtionalpotient}
\el
where $X^{\mu} = (X^0, X^i)$ are the Cartesian coordinates. The gravitational potential $U$ is defined as
\bl
U(|\vec{X}|) = \frac{G M}{|\vec{X}|}, \label{gravtionalpotient2}
\el
where $M$ is the mass of Earth and $G$ is the gravitational constant. 
The gravitomagnetic vector potential $V_i$ is defined as
\bl
V_i(|\vec{X}|) = \frac{G(\vec{J} \times \vec{X})_i}{|\vec{X}|^{3}}, \label{rotianpotient}
\el
where $\vec{J} = I \vec{\omega}$ is the angular momentum of Earth, $I$ is the moment of inertia, and $\vec{\omega}$ is the angular velocity of Earth.

From the metrics \eqref{observrefraenfram12} and \eqref{PNmetric}, one obtains (see Appendix \ref{APPad} or Ref.~\cite{Delva2017} for further details)
\begin{subequations} 
\bl
\vec{\gamma} &= \vec{a}_c - \nabla U, \label{accerlation}
\\
\vec{\omega}' &=
\vec{\omega} \left(1 + \frac{1}{2} \frac{v^2}{c^2} + \frac{U}{c^2} \right)
+ \frac{1}{2} \frac{\vec{v} \times \vec{\gamma}}{c^2}
- \frac{1}{c^2} \left( \frac{3}{2} \vec{v} \times \nabla U
+ 2 \nabla \times \vec{V} \right),\label{rotation}
\el
\end{subequations} 
where $\vec{a}_c\equiv c^2 d^2 \vec{R}/d (X^0)^2$ 
and $\vec{v}\equiv c  d \vec{R}/d X^0$
are the centripetal acceleration and the velocity of the origin of the laboratory, respectively, where $\vec{R}$ is the position vector pointing from the geocenter to the origin of the laboratory, and its length $R$ equals the Earth's radius. The gravitational potential $U = U(\vec{R})$ and the gravitomagnetic vector potential $\vec{V} = \vec{V}(\vec{R})$ are evaluated at the origin of the laboratory. The acceleration \eqref{accerlation} is kept to order $c^{0}$ since higher order terms $O(c^{-2})$ would contribute to time delays of order $O(c^{-5})$, as shown in the next section, which is beyond the scope of this paper. The angular velocity \eqref{rotation} matches the expression in Ref.~\cite{Delva2017}, but differs from that in Ref.~\cite{Bosi2009}, where couplings between $\vec{\omega}$ and $v^2, U$ were overlooked. The first term of \eqref{rotation} is referred to as the Sagnac (SA) term, the second as the Thomas precession (TP) term, the third as the Geodetic (Geo) term, and the fourth as the Lense-Thirring (LT) term \cite{Ciufolini}. Note that the vectors in \eqref{accerlation} and \eqref{rotation} are independent of the choice of basis vectors. However, to calculate gravitational effects in the laboratory, it is convenient to express them in terms of the laboratory reference frame basis vectors, see Appendix \ref{Appcoordinate} for details.

\section{Redshift and frame dragging effects in coincidence probability}\label{homdae}

In this section, we present two distinct approaches to investigate how the redshift and frame dragging effects caused by Earth's gravitational field influence the coincidence probability.
The first approach takes the particle perspective, where the two-photon state in the gravitational field is affected by relativistic time delays. This method is often used in studies of HOM experiments \cite{Brushi20141,Brushi20142,Kish2022,Restuccia2019}. The second approach takes the wave perspective, where the two-photon state is affected by phase shifts caused by the gravitational field \cite{Caiyiqi,Leo,Canad,Jerzy}. As we shall see, these two approaches lead to different coincidence probabilities.

\subsection{Particle perspective}\label{appraoch1}

From the particle perspective, light is considered a massless particle, and its interaction with the gravitational field is localized.  
In Minkowski spacetime, the coincidence probability \eqref{eq27} depends on the difference in the travel times of the two photons.  
When  considering  the gravitational field of Earth, the photon travel time undergoes an additional delay, thereby altering the coincidence probability. From the particle perspective, calculating this relativistic time delay is straightforward and only involves integrating along a photon’s null trajectory.  
In what follows, we shall compute the relativistic time delay up to second order in the coordinates $x^{\mu}$.

The null interval of a photon in the laboratory reference frame is  
\bl  
0 = g_{\mu\nu} dx^{\mu} dx^{\nu}, \label{lightinterval}  
\el  
where the metric $g_{\mu\nu}$ is given by \eqref{observrefraenfram12}.  
The infinitesimal time interval  
$dt$ can be obtained from equation \eqref{lightinterval}   
\bl  
dt = \frac{1}{c} \Bigg(-\frac{g_{0 i}}{g_{00}} dx^{i}  
+ \frac{1}{\sqrt{-g_{00}}} \sqrt{\left(g_{i j} - \frac{g_{0 i} g_{0 j}}{g_{00}}\right) dx^{i} dx^{j}}\Bigg), \label{infinintimedelay}  
\el  
where the solution with the plus sign before the square root, corresponding to the future direction,  
has been selected \cite{Landau}. Expanding $g_{\mu\nu}$ as $(\eta_{\mu\nu} + h_{\mu\nu})$ in \eqref{infinintimedelay} yields
\begin{subequations}
\bl
d\bar{t} &= \frac{1}{c} dl, \label{0thordert1} \\
dt^{(1)} &= \frac{1}{c} \left( \frac{1}{2} h_{00} dl + h_{0 i} dx^{i} + \frac{1}{2} \frac{c}{\omega} h_{ij} k^{i} dx^{j} \right), \label{1stordert1} \\
dt^{(2)} &= \frac{1}{c} \left( \frac{3}{8} h_{00}^2 dl + h_{0 i} h_{00} dx^{i} + \frac{1}{2} \frac{c}{\omega} h_{0 i} h_{0 j} k^{i} dx^{j} + \frac{1}{4} \frac{c}{\omega} h_{ij} h_{00} k^{i} dx^{j} - \frac{1}{8} \frac{c^{3}}{\omega^{3}} h_{ij} h_{mn} k^{i} k^{j} k^{n} dx^{m} \right), \label{2ndordert1}
\el
\end{subequations}
where $dl \equiv \sqrt{\delta_{ij} dx^{i} dx^{j}}$, $k^{i} \equiv \frac{\omega}{c} \frac{dx^{i}}{dl}$, and $\bar{t}$, $t^{(1)}$, and $t^{(2)}$ are the zeroth, first, and second order times expanded in terms of the metric perturbation $h_{\mu\nu}$. In the following, the times \eqref{0thordert1}, \eqref{1stordert1}, and \eqref{2ndordert1} will be expanded in terms of the coordinates $x^{\nu}$ up to second order. These intermediate expressions are written down for convenient comparison with those from the wave perspective.

Plugging \eqref{observrefraenfram12} into 
\eqref{0thordert1}, \eqref{1stordert1}, and \eqref{2ndordert1} and integrating along the light path ${\cal O}$, 
the travel time of a photon can be written as follows
\bl
t[{\cal O}]&=\bar{t}[{\cal O}]+\delta t[{\cal O}]
+\delta^{(2)} t[{\cal O}],\label{totaltimedelayto4thntheoorder}
\el
where 
\begin{subequations} \bl
\bar{t}[\cal O]&=\frac{1}{c}\int_{\cal O} dl,\label{0thapp11}
\\
\delta t [\cal O]&=\frac{1}{c^2}\int_{\cal O} 
(\vec{\omega}^{\prime} \times \vec{x})d\vec{x}-
\frac{1}{c^3}\int_{\cal O} 
(\vec{\gamma}\cdot \vec{x}) dl
,\label{0thapp21}
\\
\delta^{(2)} t [\cal O]&=-\frac{1}{c}\int_{\cal O} 
\Big[\frac{2}{c^3}(\vec{\gamma}\cdot \vec{x}) (\vec{\omega}^{\prime} \times \vec{x}) 
+ 
\frac{2}{3} R_{0lim}x^lx^m\Big]dx^i\nn
\\
&~~~~+
\frac{1}{c}\int_{\cal O}\Big[
\frac{1}{c^4} (\vec{\gamma}\cdot \vec{x})^2
-\frac{1}{2c^2} (\vec{\omega}'\cdot \vec{x})^2
+\frac{1}{2c^2}\omega'^2 \vec{x}^2-\frac12R_{0l0m} x^{l}x^m\nn
\\
&~~~~-\frac12 \frac{c^2}{\omega^2}R_{iljm}x^lx^m k^{i}  
k^{j}
+\frac12\frac{1}{\omega^2}k_{i}k_{j}(\vec{\omega}^{\prime} \times
\vec{x})^i(\vec{\omega}^{\prime} \times \vec{x})^j\Big]dl,\label{dede1}
\el
\end{subequations} 
where $\bar{t}[{\cal O}]$ is the travel time of a photon along the trajectory ${\cal O}$ in Minkowski spacetime, with ${\cal O}$ denoting ${\cal C}$ or ${\cal CC}$, as shown in Figs.~1(a) and 1(b).
$\delta t$, which is contributed solely by \eqref{1stordert1}, represents the first order relativistic time delay in terms of the coordinates $x^\mu$.  
The first and second terms of $\delta t$ correspond to the $0i$ and $00$ components of the metric \eqref{observrefraenfram12}, representing the ``frame dragging effect" and the ``redshift effect", respectively. $\delta^{(2)} t$, contributed by both \eqref{1stordert1} and \eqref{2ndordert1}, represents the second order relativistic time delay in terms of the coordinates $x^\mu$.  
The terms $h_{ij} h_{0n} \sim O(x^{3})$ and $h_{ij} h_{mn} \sim O(x^{4})$ in \eqref{2ndordert1} have been neglected in \eqref{dede1} due to their higher order contributions.

Consider a pair of photons passing through two paths of equal length, with one traveling counterclockwise along path ${\cal CC}$ and the other traveling clockwise along path ${\cal C}$.  
From the particle perspective, the gravitational field affects the travel time only through relativistic time delays. Thus, the two-photon state can be derived by making the following substitutions in \eqref{generaleq1}
\bl
t_1 &\rightarrow \bar{t}[{\cal CC}] + \delta t[{\cal CC}] + \delta^{(2)} t[{\cal CC}] \equiv \bar{t}_1 + \delta t_1 + \delta^{(2)} t_1, \label{eq388} \\
t_2 &\rightarrow \bar{t}[{\cal C}] + \delta t[{\cal C}] + \delta^{(2)} t[{\cal C}] \equiv \bar{t}_2 + \delta t_2 + \delta^{(2)} t_2. \label{eq399}
\el
Following the same procedure shown in Sec. \ref{HOMinter}, the coincidence probability can be directly written as
\bl
P^{c}(t_1 - t_2) = P^{c}(\delta t_1 - \delta t_2 + \delta^{(2)} t_1 - \delta^{(2)} t_2), \label{4666}
\el
where the Minkowski times $\bar{t}_1$ and $\bar{t}_2$, spent by photons traveling along paths ${\cal CC}$ and ${\cal C}$ respectively, are equal and hence cancel out in \eqref{4666}.
To observe the full interference pattern and detect the relativistic time delay with higher accuracy \cite{Restuccia2019}, an optical coupler can be placed in either path ${\cal CC}$ or ${\cal C}$ in the experiment, which introduces an additional time delay ${\cal T}$ by fine-tuning the position of the coupler \cite{Restuccia2019}.  
In what follows, we assume the coupler is placed in path ${\cal CC}$, consistent with Ref. \cite{Restuccia2019}. Therefore, the coincidence probability becomes
\bl
P^{c}({\cal T} + \delta t_1 - \delta t_2 + \delta^{(2)} t_1 - \delta^{(2)} t_2). \label{eqaiotn40}
\el
Note that if the coupler is placed in path ${\cal C}$, the coincidence probability becomes
\bl
P^{c}(\delta t_1 - \delta t_2 + \delta^{(2)} t_1 - \delta^{(2)} t_2 - {\cal T}).
\el

\subsection{Wave perspective}\label{appraoch2}

From the wave perspective, light is considered a wave, and the interaction 
with the gravitational field is nonlocal. To explicitly {exhibit} 
the differences between the 
particle and wave perspectives, we shall calculate the phase shift (and the corresponding 
effective time delay) of a photon up to second order 
by solving the Klein-Gordon equation in the laboratory reference frame, 
which {has not} been discussed in Refs. \cite{Brandy2021,Caiyiqi,Leo,Canad,Jerzy}. 
Furthermore, we will calculate the coincidence probability  
using the two-photon state constructed from
the effective time delay rather than from the relativistic time delay.

Since in this work we mainly consider the phase shift as the dominant factor affecting the indistinguishability of the two photons, and the polarization of photons does not affect the coincidence probability, for simplicity the electromagnetic field can be approximated by a massless, minimally-coupling scalar field in curved spacetime  \cite{Fulling1989, Davies1982, Wald1994, Parker2009, BU2020},
\bl
\nabla_{\alpha}\nabla^{\alpha}\phi(x)=0,\label{eomp}
\el
where \(\nabla_{\alpha}\) is the covariant derivative, which reduces to the ordinary derivative \(\partial_{\alpha}\) in Minkowski spacetime.  
In curved spacetime, the equations of motion for the two physical transverse modes of the Maxwell field are equivalent to {those of} the conformally-coupling massless scalar field \cite{YeZhang2,YeZhang22,YeZhang22ee,Zhang1,Zhang2,YeZhang} and differ from \eqref{eomp}.  
Generally, the propagation of a scalar field can be studied by adding the term \(\xi R \phi(x)\), where \(\xi\) is a coupling constant, to the equation of motion \eqref{eomp}. This aspect will be investigated in follow-up research.

We shall quantize the scalar field below.
Write
\bl
\phi(x)&=\int \frac{d^3\vec{k}}{(2\pi)^3} [a_{\vec{k}} u_{\vec{k}} + a_{\vec{k}}^{\dag}u_{\vec{k}}^{*}], \label{scalrafield53}
\el
where $u_{k}$ is the mode function, and $a_{\vec{k}}$ and $a_{\vec{k}}^{\dag}$
are the annihilation and creation
operators satisfying the commutation relation $[a_{\vec{k}},
a_{\vec{k}}^{\dag}]=\delta^{(3)}(\vec{k}-\vec{k})$. Imposing 
the equal time commutation relation $[\phi(t,\vec{x}),\pi(t, \vec{x}')]=i\delta^{(3)}(\vec{x}-\vec{x}')$, where $\pi\equiv\partial\phi/\partial t$ is the conjugate 
momentum, yields  
\bl
u_{\vec{k}}\dot{u}_{\vec{k}}^*
-u_{\vec{k}}^*\dot{u}_{\vec{k}}=i,\label{wronskian}
\el
which is the Wronskian condition and is useful for normalizing the mode function. In the geometric optics approximation \cite{MTW,Landau}, the mode function is
\bl
u_{\vec{k}}=\alpha_{\vec{k}}' e^{iS_{\vec{k}}'}, \label{eq54}
\el
where $S_{\vec{k}}'$ is the eikonal, and \(\alpha_{\vec{k}}'\) is the slowly varying envelope with respect to variations in the eikonal. We shall solve the eikonal order by order.  
Plugging \eqref{scalrafield53} and \eqref{eq54} into \eqref{eomp} leads to the eikonal equation \cite{MTW,Canad,Jerzy},
\bl
g^{\mu\nu}\nabla_{\mu} S'_{\vec{k}}
\cdot\nabla_{\nu} S'_{\vec{k}}=0,\label{eqfjrieo}
\el
where the slowly varying 
envelope approximation 
$\nabla_{\alpha}\nabla^{\alpha} \alpha_{\vec{k}}'\simeq0$ has been used \cite{Brandy2021}.
\eqref{eq54} is referred to as the optics approximation because 
the form of \eqref{eqfjrieo} is identical to the null interval equation 
\eqref{lightinterval}.
Expanding the eikonal in terms of the metric perturbation $h_{\mu\nu}$ up to the second order yields 
\bl
S'_{\vec{k}}=\bar{S}_{\vec{k}}+ S_{\vec{k}}^{(1)}+ S_{\vec{k}}^{(2)},\label{expandingto2ndorder}
\el
where $\bar{S}_{\vec{k}}, S_{\vec{k}}^{(1)},S_{\vec{k}}^{(2)}$ are the 
zeroth, first and second order phases, respectively.
By plugging \eqref{expandingto2ndorder} into \eqref{eqfjrieo} and separating terms according to their order, we obtain
\begin{subequations} 
\bl
0&=\eta^{\alpha\beta}\partial_{\alpha} \bar{S}_{\vec{k}}
\partial_{\beta} \bar{S}_{\vec{k}},\label{eqss0th}
\\
0&=2\eta^{\alpha\beta}\partial_{\alpha}\bar{S}_{\vec{k}}
\partial_{\beta} S^{(1)}_{\vec{k}}-h^{\alpha\beta}
\partial_{\alpha}\bar{S}_{\vec{k}}\partial_{\beta}\bar{S}_{\vec{k}},\label{eqss1st}
\\
0&=2\eta^{\alpha\beta}\partial_{\alpha}\bar{S}_{\vec{k}} \partial_{\beta}\delta^{(2)} S_{\vec{k}}
+\eta^{\alpha\beta}\partial_{\alpha} S^{(1)}_{\vec{k}} \partial_{\beta}S^{(1)}_{\vec{k}}
-2h^{\alpha\beta}\partial_{\alpha}\bar{S}_{\vec{k}}\partial_{\beta} S^{(1)}_{\vec{k}}
+h^{\alpha\sigma}h_{\sigma}^{~\beta}\partial_{\alpha}\bar{S}_{\vec{k}}\partial_{\beta} 
\bar{S}_{\vec{k}},\label{eqss2nd}
\el 
\end{subequations} 
where the covariant derivative has been replaced by the ordinary derivative, and  
$g^{\mu\nu}=\eta^{\mu\nu}-h^{\mu\nu}+h^{\mu\alpha}h_{\alpha}^{~\nu}$ is used.
This expression is derived from the relations 
$g_{\mu\alpha}g^{\alpha\nu}=\delta_{\mu}^{~\nu}$ and 
$g_{\mu\alpha}=\eta_{\mu\nu}+h_{\mu\nu}$. 
Eqs. \eqref{eqss0th}, \eqref{eqss1st}, and \eqref{eqss2nd} can be written  as 
\begin{subequations} 
\bl
k_{\alpha}k^{\alpha}&=0,\label{improkequ}
\\
\partial_{\alpha}S^{(1)}_{\vec{k}}&=\frac12
h_{\alpha\rho}k^{\rho},\label{sohdieuwfh1}
\\
\partial_{\alpha} S^{(2)}_{\vec{k}}&=-\frac18 h_{\rho\alpha}
h^{\rho\sigma}k_{\sigma},\label{sohdieuwfh2}
\el
\end{subequations} 
where $k_\alpha \equiv \partial_\alpha S_{\vec{k}} = (-\omega/c, k^i)$ is a null vector in Minkowski spacetime.  
Thus, $d S_{\vec{k}}$, $d S^{(1)}_{\vec{k}}$, and $d S^{(2)}_{\vec{k}}$ are 
\begin{subequations} 
\bl
d S_{\vec{k}}&=-\frac{\omega}{c} dx^0 + k_i dx^i,\label{dewfew0}
\\
d S^{(1)}_{\vec{k}}&=\frac{\omega}{c} \Big(\frac12 h_{00}dl+h_{0 i} d x^{i} 
+\frac12\frac{c}{\omega}h_{ij}k^idx^j 
\Big),\label{dewfew}
\\
dS^{(2)}_{\vec{k}}&
=\frac18\frac{\omega}{c}\Big(h_{0 0}^2dl
-h_{j 0}
h^{j}_{~0} d l
+2\frac{c}{\omega}h_{0 i}
h_{00}k^{i} d l
-2\frac{c}{\omega}h_{j i}
h^{j}_{~0}k^{i} d l
+\frac{c^2}{\omega^2}h_{0 j}
h_{0i}k^{i}k^{j} d l
-\frac{c^2}{\omega^2}h_{k j}
h^{k}_{~i}k^{i}k^{j} d l\Big).\label{dewfew2}
\el
\end{subequations} 
Comparing \eqref{dewfew} 
with \eqref{1stordert1}, and 
\eqref{dewfew2} with \eqref{2ndordert1}, one finds  
\begin{subequations} 
\bl
d t^{(1)}&=- d S^{(1)}_{\vec{k}}/(ck_0)\neq d S^{(1)}_{\vec{k}}/(c k_0),
\\
d t^{(2)}&\neq d S^{(2)}_{\vec{k}}/(c k_0),
\el
\end{subequations} 
where $d t^{(1)}$ differs from $d S^{(1)}_{\vec{k}}/(c k_0)$ by a minus sign, and  
$d t^{(2)}$ even differs from  
$d S_{\vec{k}}^{(2)}/(c k_0)$ by several terms,  
which shows that the time delay cannot  
be trivially obtained by dividing the  
phase shift by $c k_0$. The definition $d t \equiv d S_{\vec{k}}/(c k_0)$ is  
incorrect because the phase consists of information from both the  
temporal and {spatial} parts, as can be clearly seen in \eqref{dewfew0}.  
In Ref.~\cite{Brandy2021}, the first order time delay obtained using  
$d t^{(1)} = d S^{(1)}_{\vec{k}}/(c k_0)$ is the same as \eqref{1stordert1}. However, this  
agreement is coincidental because (2.12) in Ref.~\cite{Brandy2021}  
differs from \eqref{sohdieuwfh1} by a minus sign.  
In fact, as we have shown in Appendix~\ref{Appcontoti},  
the phase shift and the time delay are related by  
a non-trivial coefficient vector $\Lambda^{0}_{~\alpha}$.

The phases at various orders can be obtained by integrating \eqref{improkequ}, \eqref{sohdieuwfh1}, and \eqref{sohdieuwfh2} over the path $\mathcal{O}$, as follows
\begin{subequations} 
\bl
\bar{S}_{\vec{k}}[{\cal O}]&=\int_{{\cal O}} k_{\alpha}dx^{\alpha},\label{S0}
\\
S^{(1)}_{\vec{k}}[{\cal O}]&=\frac12
\int_{\cal O} h_{\alpha\rho}k^{\rho} dx^{\alpha},\label{S1}
\\
S^{(2)}_{\vec{k}}[{\cal O}]&=-\frac18 \int_{\cal O}h_{\rho\alpha}
h^{\rho\sigma}k_{\sigma}dx^{\alpha},\label{S2}
\el
\end{subequations} 
the first and second order eikonals, $S^{(1)}_{\vec{k}}$ and $S^{(2)}_{\vec{k}}$ are proportional to time since the metric perturbation $h_{\mu\nu}$ is time-independent.  
The Wronskian condition \eqref{wronskian} can be satisfied by adjusting the coefficient of $e^{iS_{\vec{k}}'}$.  
Thus, canonical quantization can be performed in a weak gravitational field.  
The normalized mode function can be obtained by replacing $\alpha_{\vec{k}}'$ in \eqref{eq54} with 
${\cal N}_{\alpha_{\vec{k}}'}$, 
\bl
u_{\vec{k}}
&=
{\cal N}_{\alpha_{\vec{k}}'} 
e^{-i \omega\big( \bar{t}+\delta T+\delta^{(2)} T\big)}e^{ik \int_{\cal O}dl},\label{scalrafield53333}
\el
where the relations $k_0=-\omega/c$, $cdt=dl$, $\vec{k}\cdot\hat{k}=\vec{k}\cdot d\vec{x}/dl=k$, and the definition \eqref{0thapp11} have been used. 
$\delta T$ and $\delta^{(2)} T$ denote the 1st and 2nd order effective time delays, which are 
defined as follows 
 \bl
\delta T[{\cal O}]=-\frac{1}{c^2}\int_{\cal O}
(\vec{\omega}^{\prime} \times \vec{x})^id x^{i}
+\frac1{c^3}\int_{\cal O}
(\vec{\gamma}\cdot \vec{x})d l\label{phasetotime11}
\el
and 
\bl
\delta^{(2)} T[{\cal O}]&=\frac{1}{2c}\int_{\cal O}
[
\frac{1}{c^2} (\vec{\omega}'\cdot \vec{x})^2
-\frac{1}{c^2}\omega'^2 \vec{x}^2
+R_{0l0m} x^{l}x^m]d l+\frac{1}{c}\frac23\int_{\cal O}
 R_{0lim}x^lx^m d x^{i}+
\frac1{2\omega}\int_{\cal O}
R_{ilnm}x^lx^m k^{i} dx^n\nn
\\
&
-\frac{1}{2c^4}\int_{\cal O}(\vec{\gamma}\cdot \vec{x})(\vec{\omega}^{\prime} \times \vec{x})^i dx^{i}
+\frac{1}{8c^3}\int_{\cal O} (\vec{\omega}^{\prime}
\times \vec{x})_i(\vec{\omega}^{\prime} \times \vec{x})^i 
d l-
\frac{1}{8c\omega^2}\int_{\cal O} 
(\vec{\omega}^{\prime} 
\times \vec{x})_i(\vec{\omega}^{\prime} 
\times \vec{x})_j\hat{k}^i \hat{k}^jd l,\label{phasetotime22}
\el
where the metric \eqref{observrefraenfram12} has been used.  
The terms of order $O(x)$ in $(-S_{\vec{k}}^{(1)}/\omega)$ and those of order $O(x^2)$ in $(-S_{\vec{k}}^{(2)}/\omega)$  are respectively collected in \eqref{phasetotime11} and \eqref{phasetotime22}.  
By comparing \eqref{phasetotime11} and \eqref{phasetotime22} with \eqref{0thapp21} and \eqref{dede1}, one finds that the effective time delays \(\delta T\) and \(\delta^{(2)} T\) obtained from the phase shifts are not equal to the relativistic time delays, namely, $\delta t=-\delta T$ and $\delta^{(2)} t\neq\delta^{(2)} T$, which leads to different coincidence probabilities.  
In the following, we shall derive the coincidence probability using the two-photon state constructed from the phase.

In the HOM interference experiments, the trajectory and  central frequency of a photon
are determined by the unit vector 
$\hat{k}$ and $\omega=c|\vec{k}|$.
The single photon operator can be written as  
\bl
\phi_{\vec{k}}^{\dag}\equiv \int \Phi(\omega, \boldsymbol{\sigma}) a_{\vec{k}}^{\dag} u_{\vec{k}}^* d\omega 
=
{\cal N}_{\alpha_{\vec{k}}'}^{*}\int d\omega \Phi(\omega, \boldsymbol{\sigma})
a_{\boldsymbol{\sigma}}^{\dag}(\omega)
e^{-iS_{\vec{k}}'},\label{scalrafield533}
\el
where $\Phi(\omega, \boldsymbol{\sigma})$ is the spectral amplitude function 
of the quasimonochromatic photon, {\small ${\cal N}_{\alpha_{\vec{k}}'}^*$} is the 
normalization constant, and the notation $a_{\boldsymbol{\sigma}}^{\dag}(\omega)$ is 
used instead of $a_{\vec{k}}^{\dag}$, with the degree
of freedom of $\hat{k}$ being absorbed into $\boldsymbol{\sigma}$.
The single-photon state is given by 
$\phi^{\dag}_{\vec{k}}|0\rangle$, and  the two-photon 
state  is 
\bl
|\phi( S_{\vec{k}_1}',S_{\vec{k}_2}')\rangle
& \equiv \hat{\phi}^{\dag}_{\vec{k}_1}\hat{\phi}^{\dag}_{\vec{k}_2}|0\rangle
={\cal N}_{\phi}\int d\omega_1d\omega_2\Phi_1(\omega_1,\boldsymbol{\sigma}_1)\Phi_2(\omega_2,
\boldsymbol{\sigma}_1)
a_{\boldsymbol{\sigma}_1}^{\dag}(\omega_1)a_{\boldsymbol{\sigma}_2}^{\dag}(\omega_1) e^{-iS_{\vec{k}_1}'}
e^{-iS_{\vec{k}_2}'}
|0\rangle,\label{twophotonstate}
\el
where {\small ${\cal N}_{\phi}\equiv{\cal N}_{\alpha_{\vec{k}_1}'}^*
{\cal N}_{\alpha_{\vec{k}_2}'}^*$. } The state \eqref{twophotonstate} is consistent with the state \eqref{generaleq1} because the zeroth order temporal part of \( e^{-i S'_{\vec{k}_1}} \) reduces to \( e^{i \omega_1 t_1} \) in \eqref{generaleq1}, and the same holds for \( e^{-i S'_{\vec{k}_2}} \).  
\( S'_{\vec{k}_1} \) and \( S'_{\vec{k}_2} \) are the phases of the photons traveling along the paths \(\mathcal{CC}\) and \(\mathcal{C}\), respectively.  
The state \eqref{twophotonstate} describes photons that are spectrally indistinguishable and frequency-uncorrelated.  
A generalization of the state \eqref{twophotonstate} can be made by replacing \(\Phi_1(\omega_1,\boldsymbol{\sigma}_1) \Phi_2(\omega_2,\boldsymbol{\sigma}_2)\) with \(\Phi(\omega_1, \omega_2, \boldsymbol{\sigma}_1, \boldsymbol{\sigma}_2)\), which describes a frequency-correlated photon pair.  
Summing over other discrete degrees of freedom of the photons yields
\bl
|\phi( S_{\vec{k_1}}',S_{\vec{k_2}}')\rangle
&=\sum_{\boldsymbol{\sigma}_1,\boldsymbol{\sigma}_2}{\cal N}_{\phi}
\int d\omega_1d\omega_2\Phi_1(\omega_1,\omega_2,\boldsymbol{\sigma}_1,
\boldsymbol{\sigma}_2)
a_{\boldsymbol{\sigma}_1}^{\dag}(\omega_1)a_{\boldsymbol{\sigma}_2}^{\dag}(\omega_2) 
e^{i\omega_1(\bar{t}_1+\delta T_1+\delta^{(2)} T_1)}
e^{i\omega_2(\bar{t}_2+\delta T_2+\delta^{(2)} T_2)}
|0\rangle,\label{twophotonstate21}
\el
where \eqref{scalrafield53333} has been used. Note that the factor
$
e^{i \left(k_1 \int_{\cal CC} dl + k_2 \int_{\cal C} dl\right)} = e^{i \left(\frac{\omega_1}{c} \int_{\cal CC} dl + \frac{\omega_2}{c} \int_{\cal C} dl\right)}
$
is omitted in \eqref{twophotonstate21} because the two photons travel equal distances in the HOM experiment. According to \eqref{jointprperwrittingpaper}, this factor affects neither the difference \((t_1 - t_2)\) nor the coincidence probability.

The state \eqref{twophotonstate21} is similar to \eqref{generaleq1}, by following the same process in Sec. \ref{HOMinter}, one can directly write down the coincidence probability  
\bl
P^{c}({\cal T} + \delta T_1 - \delta T_2 + \delta^{(2)} T_1 - \delta^{(2)} T_2), \label{68huiwh}
\el
where \({\cal T}\) was defined in Sec. \ref{appraoch1}. The coincidence probabilities \eqref{eqaiotn40} and \eqref{68huiwh} differ even when the second order time delays are neglected. Specifically, \(\delta t\) differs from \(\delta T\) by a minus sign, resulting in the coincidence probability \(P^{c}({\cal T})\) shifting in opposite directions, as will be shown in Sec. \ref{revisit}.

The second order time delays are currently too small to be detected. We calculate them explicitly to clearly demonstrate that the relativistic and effective time delays are different. For the remainder of the paper, we consider only the first order time delays.

\section{Experimental setup }\label{expsetup}

\begin{figure}[htb]
 \centering
 \subcaptionbox{}
   {%
     \includegraphics[width = .36\linewidth]{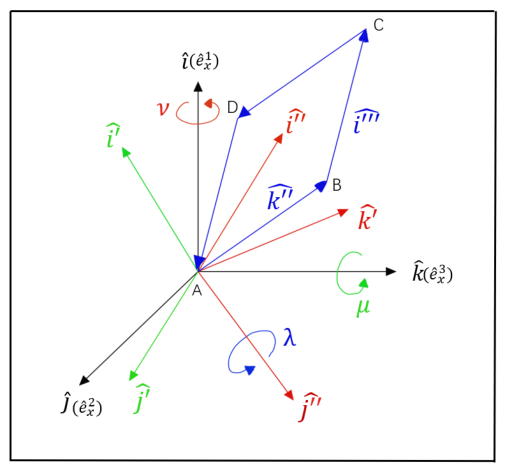}}
      \hspace{.50in}
      \subcaptionbox{}
   {%
     \includegraphics[width = .35
\linewidth]{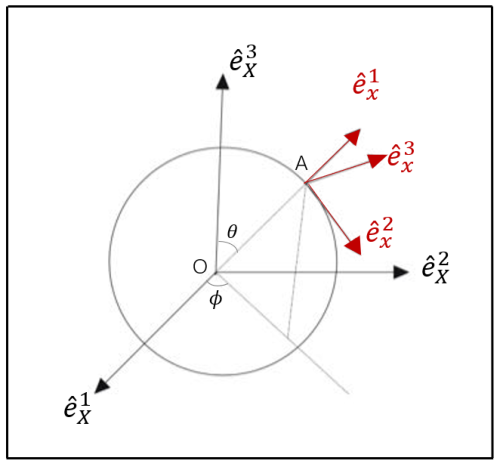}}  
\caption{  
 (a) Representation of interferometer vectors and the 
 definition of Euler angles ($\mu,\nu,\lambda$). (b) The location of the laboratory on the 
 Earth with associated basis vectors. } \label{intermorefer}
\end{figure}

In this section, we will parameterize the quantities in the interferometer, shown in Fig.\ref{intermoreferconfing}, within the laboratory coordinate system. The interferometer $ABCD$ in Fig.\ref{intermorefer} (a) corresponds to that shown in Figs. \ref{intermoreferconfing} (a) and (b).
The origin of the laboratory coordinate system is taken to coincide with that of the interferometer, denoted by $A$, shown in Fig. \ref{intermorefer}(b), whose coordinates in the geocentric coordinate system are $(R, \theta, \phi)$. The orthogonal basis vectors of the geocentric coordinate system, $\hat{e}^i_X$, and of the laboratory coordinate system, $\hat{e}^i_x$, are also shown in Fig. \ref{intermorefer}(b). Relevant quantities with respect to these basis vectors are given in Appendix \ref{Appcoordinate}. 

The interferometer is rectangular, 
with four arms defined by
\begin{subequations} 
\bl
\vec{L}_{AB}&\equiv\vec{x}_{B}-\vec{x}_{A},\label{definaBA}
\\
\vec{L}_{BC}&\equiv\vec{x}_{C}-\vec{x}_{B},\label{definaCB}
\\
\vec{L}_{CD}&=-\vec{L}_{AB},\label{definaDC}
\\
\vec{L}_{DA}&=-\vec{L}_{BC},\label{definaAD}
\el
\end{subequations} 
the adjacent arms are perpendicular to each other, 
the unit vector of each arm is denoted as $\hat{L}_{BA}$, etc. 
The areal vector is defined as 
\bl
\vec{{\cal A}}\equiv{\cal A} \hat{{\cal A}}={\cal A}
(\hat{L}_{AB}\times\hat{L}_{BC}),\label{unitarea}
\el
where ${\cal A}$ is the area of the interferometer and $\hat{{\cal A}}$ is the unit 
areal vector.

We shall express \(\hat{{\cal A}}\), \(\hat{L}_{AB}\), and \(\hat{L}_{BC}\) in terms of three Euler angles \((\mu, \nu, \lambda)\) and the basis vectors \(\hat{e}^i_x\). The Euler angles are defined as follows. Suppose the initial orthogonal unit vectors \(\hat{i}, \hat{j}, \hat{k}\) are parallel to the basis vectors \(\hat{e}_x^1, \hat{e}_x^2, \hat{e}_x^3\) of the laboratory coordinate system.
First, we rotate \(\hat{i}, \hat{j}, \hat{k}\) counterclockwise about the \(z\)-axis by an angle \(\mu\), represented by \(R_z(\mu)\). The new orthogonal unit vectors are labeled \(\hat{i}', \hat{j}', \hat{k}\). Next, we rotate the vectors \(\hat{i}', \hat{j}', \hat{k}\) counterclockwise about the \(x\)-axis by an angle \(\nu\), represented by \(R_x(\nu)\). The new orthogonal unit vectors are labeled \(\hat{i}'', \hat{j}'', \hat{k}'\). Then, we rotate the vectors \(\hat{i}'', \hat{j}'', \hat{k}'\) counterclockwise about \(\hat{j}''\) by an angle \(\lambda\), represented by
$R_{j''}(\lambda) = R_x(\nu) R_z(\mu) R_y(\lambda) R_z^{-1}(\mu) R_x^{-1}(\nu)$.
This results in the unit vectors \(\hat{i}''', \hat{j}'', \hat{k}''\). The rotation matrix \(R_h(\alpha)\) represents a rotation by an angle \(\alpha\) about the \(h\)-axis. \(R_h^{-1}(\alpha)\) denotes the inverse of \(R_h(\alpha)\). These rotation matrices can be found in Ref. \cite{sakurai2020}. These steps are shown in Fig. \ref{intermorefer}(a).
The matrix of these rotations is
\bl
R(\lambda,\mu,\nu)\equiv R_{j''}(\lambda)R_{x}(\nu)R_{z}(\mu)=R_{x}(\nu)R_{z}(\mu)R_{y}(\lambda).\label{generalrotaitonmetric}
\el
With the help of \eqref{generalrotaitonmetric}, one has  
\begin{subequations} {\small\bl
\hat{{\cal A}}&\equiv R(\lambda,\mu,\nu) \hat{j}=-\sin\mu \hat{e}_x^1+\cos\mu\cos\nu \hat{e}_x^2+\cos\mu\sin\nu \hat{e}_x^3,\label{uniA}
\\
\hat{L}_{AB}&\equiv R(\lambda,\mu,\nu)\hat{k}=\cos\mu\sin\lambda\hat{e}_x^1+(\cos\nu\sin\lambda\sin\mu-\cos\lambda\sin\nu)\hat{e}_x^2
+(\cos\lambda\cos\nu+\sin\lambda\sin\mu\sin\nu)\hat{e}_x^3,\label{uniCB}
\\
\hat{L}_{BC}&\equiv R(\lambda,\mu,\nu)\hat{i}=\cos\lambda\cos\mu\hat{e}_x^1+(\cos\lambda\cos\nu\sin\mu+\sin\lambda\sin\nu)\hat{e}_x^2
+(-\cos\nu\sin\lambda+\cos\lambda\sin\mu\sin\nu)\hat{e}_x^3.\label{uniBA}
\el}
\end{subequations} 
Here, \(R(\lambda,\mu,\nu)\) is the rotation matrix and should not be confused with the radius of the Earth, \(R\). The set of unit vectors can be oriented in an arbitrary direction, which generalizes the case in Ref. \cite{Brandy2021}. By setting the angles \(\nu=0\) and \(\mu=-(\frac{\pi}{2} + \alpha)\), \eqref{uniA} becomes
\bl
\hat{\cal A}=(\hat{e}_{x}^1\cos\alpha-\hat{e}_{x}^2\sin\alpha),
\label{Brandyschoseuni}
\el
where \(\alpha\) is the angle between \(\hat{e}_{x}^1\) and the areal vector \(\hat{\cal A}\).  
The unit areal vector given in \eqref{Brandyschoseuni} is the same as that in Ref. \cite{Brandy2021}.  
Setting the angles \(\nu=0\), \(\lambda=\alpha\), and \(\mu=\beta\) in \eqref{uniCB} and \eqref{uniBA} yields
\begin{subequations} 
\begin{align}
\hat{L}_{AB} &= -\hat{L}_{BA} = \cos\beta \sin\alpha\, \hat{e}_x^1 + \sin\alpha \sin\beta\, \hat{e}_x^2 + \cos\alpha\, \hat{e}_x^3, \label{LBAredice} \\
\hat{L}_{BC} &= -\hat{L}_{CB} = \cos\alpha \cos\beta\, \hat{e}_x^1 + \cos\alpha \sin\beta\, \hat{e}_x^2 - \sin\alpha\, \hat{e}_x^3, \label{LBCredice}
\end{align}
\end{subequations} 
where \(\hat{L}_{AB}\) and \(\hat{L}_{BC}\) are the unit arm vectors used in Ref. \cite{Brandy2021}.  
The angle \(\alpha\) in \eqref{Brandyschoseuni}, \eqref{LBAredice}, and \eqref{LBCredice} does not represent the same angle as in Ref. \cite{Brandy2021}, where the authors did not distinguish \(\alpha\) in these formulas.  
We shall use \eqref{uniA}, \eqref{uniCB}, and \eqref{uniBA} in the following sections.

\section{Time delays}\label{analysetimedelay}

In this section, we calculate the first order relativistic and the effective time delays for two scenarios, and determine how many loops photons must travel in the interferometer to achieve the current experimental accuracy. The additional time delay \({\cal T}\), which is controlled by the experimentalist \cite{Restuccia2019}, does not involve gravity, and the second order time delays are too small to be measured in the current experiment. Therefore, they will not be considered. For simplicity of notation, we only present the effective time delay \((\delta T_1 - \delta T_2)\), since it can explain the turntable HOM experiment in Sec.\ref{revisit}. The first order relativistic time delay \((\delta t_1 - \delta t_2)\) can be obtained by taking the negative of the effective one, this should be kept in mind when comparing with the time delays in Refs. \cite{Brandy2021,Bosi2009}.

\subsection{Paths $ABCDA$
and $ADCBA$: frame dragging time delay}\label{frg}

Consider the interferometer configuration described in Fig. \ref{intermoreferconfing}(a) and Fig. \ref{intermorefer}(a). Two photons travel along the counterclockwise path \(\mathcal{CC}: ABCDA\) and the clockwise path \(\mathcal{C}: ADCBA\), respectively.  According to \eqref{phasetotime11}, the difference between the first order effective time delays of paths \(\mathcal{CC}\) and \(\mathcal{C}\) is given by
\bl
(\delta T_1-\delta T_2)_{FD}
&\equiv-\frac{1}{c^2}\int_{\cal CC} d x^{i}
 (\vec{\omega}^{\prime} \times \vec{x})^i+
 \frac{1}{c^2}\int_{\cal C} d x^{i}(\vec{\omega}^{\prime} \times \vec{x})^i\label{leadingorfrt}
 \\
 &=
-\frac{4\vec{\cal A}\cdot \vec{\omega}}{c^2}
-\frac{2\vec{\cal A}\cdot \vec{\omega}}{c^4}\big((\omega R)^2
\sin^2\theta-\frac{2G M}{R}\big)\nn
\\
&~~~-\frac{2\omega\vec{\cal A}}{c^4} 
\cdot\Big[-\frac{G M}{ R}\hat{e}_{x}^2
+R^2\omega^2\sin\theta(\cos\theta \hat{e}_{x}^1
- \sin\theta\hat{e}_{x}^2)\Big]\sin\theta \nn
\\
&~~~+\frac{6\omega\vec{\cal A}\cdot\hat{e}_{x}^2}{c^4} \frac{G M}{R}\sin\theta
-\frac{4IG\omega}{c^4R^3}\vec{{\cal A}}
\cdot(-2\cos\theta \hat{e}_{x}^1-\sin\theta 
\hat{e}_{x}^2),\label{aboveformula}
\el
where \eqref{1222}, \eqref{leadingoreerThom}, \eqref{leadingoreerGeo}, and \eqref{leadingoreerLT} have been used.  The subscript ``FD" denotes the frame dragging effect.  
The time delay \eqref{aboveformula} consists of five parts: the leading order Sagnac effect, the next-to-leading order Sagnac effect, the Thomas precession effect, the Geodetic effect, and the Lense-Thirring effect \cite{Ciufolini}.  Since the redshift effect,  
$\frac{1}{c^3} \left[ \int_{\cal CC} - \int_{\cal C} \right] (\vec{\gamma} \cdot \vec{x})\, dl$,  
vanishes when the paths are \({\cal CC}: ABCDA\) and \({\cal C}: ADCBA\), only the frame dragging effect survives in \eqref{leadingorfrt}.  To make the redshift effect appear, one could choose the paths \({\cal CC}: CBA\) and \({\cal C}: CDA\), which will be shown in the next subsection.

The time delay \eqref{aboveformula} depends on the area enclosed by the counterclockwise light trajectory and the orientation of the interference plane.  In what follows, the ratio of the time delays to the area of the interferometer will be used to estimate the magnitude of the various effects \cite{Brandy2021}. Using the relevant physical quantities of the Earth given in Appendix \ref{Appanlase}, the order of magnitude of the leading order Sagnac effect is
\bl
-\frac{4\hat{\cal A}\cdot \vec{\omega}}{c^2}
\approx  10^{-15} ~\text{s/km}^2,
\label{timedesagc1}
\el
the  next-to-leading order Sagnac effect by the gravitational acceleration is 
\bl
\frac{2\hat{\cal A}\cdot \vec{\omega}}{c^4 }
\frac{2G M}{R}  \approx   10^{-24}~\text{s/km}^2,
\label{timedesagc2}
\el
and by the centripetal acceleration is 
\bl
-\frac{2\hat{\cal A}\cdot \vec{\omega}}{c^4 }
(\omega R)^2 \sin^2\theta 
\approx  10^{-27} ~\text{s/km}^2.
\label{timedesagc3}
\el
The order of magnitude of the Thomas precession by 
the gravitational acceleration
is
\bl
\frac{2\omega\hat{\cal A}}{c^4 }
\cdot\big(\frac{G M}{ R}\hat{e}_{x}^2 \big)\sin\theta
\approx  10^{-24} ~\text{s/km}^2,\label{timedeTM1}
\el
and by  the centripetal acceleration is
\bl
-\frac{2\omega\hat{\cal A}}{c^4  }
\cdot\Big( R^2\omega^2\sin\theta(\cos\theta \hat{e}_{x}^1
- \sin\theta\hat{e}_{x}^2)\Big)\sin\theta
\approx  10^{-27} ~\text{s/km}^2.\label{timedeTM2}
\el
The orders of magnitude of the Geodetic effect  is 
\bl
-\frac{6\omega\hat{\cal A}\cdot\hat{e}_{x}^2}{c^4} \frac{G M}{R}\sin\theta
\approx  10^{-24}~\text{s/km}^2.\label{timedeTMm}
\el
The orders of magnitude  of the  Lense-Thirring effect is 
\bl
\frac{4IG\omega}{c^4R^3}\hat{{\cal A}}
\cdot(2\cos\theta \hat{e}_{x}^1+\sin\theta 
\hat{e}_{x}^2)
\approx 10^{-24}~\text{s/km}^2.\label{timedeLT}
\el
The effective time delays in \eqref{timedesagc3}  and \eqref{timedeTM2}
due to the centripetal acceleration are $\sim 10^{-27}$s
and are dropped in the following, 
the remains of \eqref{aboveformula} 
are (using \eqref{Brandyschoseuni} and dropping terms $\sim \omega^3$)
\bl
(\delta T_1-\delta T_2)_{FD}
&=-\frac{4\omega{\cal A}}{c^2}\cos(\theta-\alpha)
-\frac{4\omega{\cal A}}{c^4}
  \frac{2G M}{ R}\sin\alpha\sin\theta\nn
\\
&~~~~~~~~~~~~~~~-\frac{4\omega{\cal A}}{c^4} \frac{I G}{R^3}
(-2\cos\alpha \cos\theta
+\sin\alpha\sin\theta )
+\frac{4\omega{\cal A}}{c^4} \frac{G M}{R}\cos(\theta-\alpha),
\label{finalexample}
\el
where \(\nu=0\) and \(\mu= - \left(\frac{\pi}{2} + \alpha\right)\) have been used.  
\eqref{finalexample} is consistent with the results in Refs. \cite{Brandy2021,Bosi2009}, except for the last term, $\frac{4 \omega \mathcal{A}}{c^4} \left( \frac{G M}{R} \cos(\theta - \alpha) \right)$.  
This term arises from the next-to-leading order Sagnac effect, which is of the same order of magnitude as the second and third terms, and therefore should be taken into consideration, as mentioned below \eqref{rotation}.  The second term in \eqref{finalexample} results from both the Geodetic effect and the Thomas precession effect caused by gravitational acceleration.

\begin{figure}[htb]
 \centering
 \subcaptionbox{}
   {%
     \includegraphics[width = .4\linewidth]{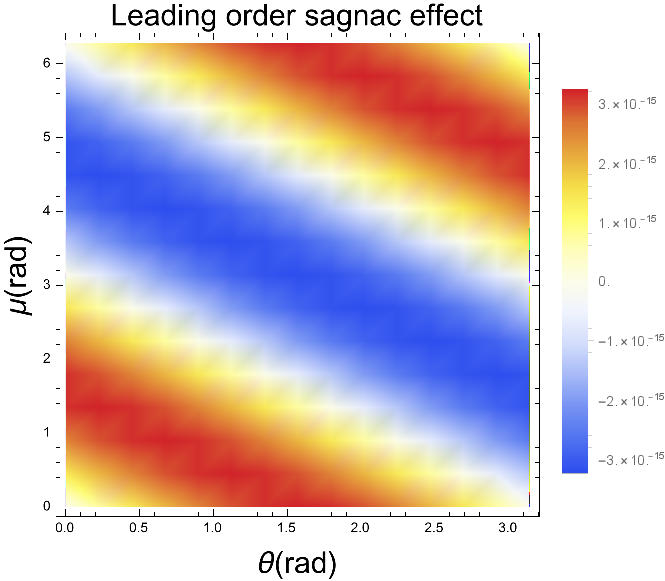}}
     \hspace{.5in}
 \subcaptionbox{}
   {%
     \includegraphics[width = .4 
\linewidth]{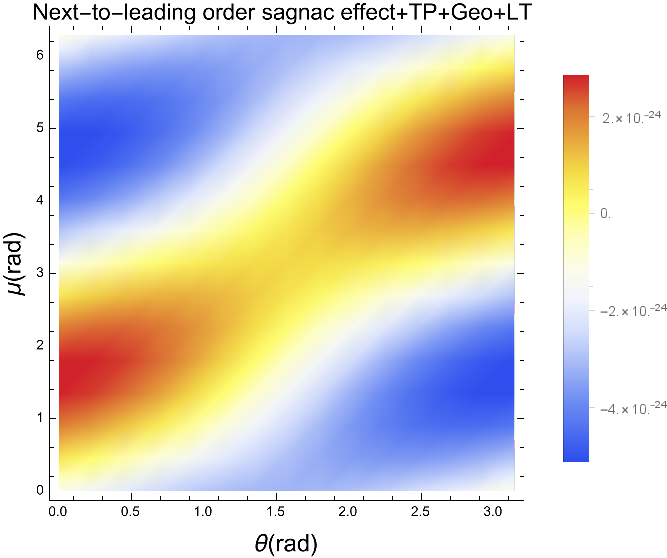}} 
\caption{ (a) Time delay due to the leading order Sagnac effect. (b) 
Time delay due to the combined effects of the next-to-leading order Sagnac effect, Thomas precession effect, Geodetic effect and  Lense-Thirring effect. Interferometer area: $1 \text{km}^2$.  Color: time delay, unit: s.} \label{fig}
\end{figure}

For illustration, substituting \eqref{uniA} into \eqref{aboveformula}, setting the Euler angle \(\nu = 0\), and ignoring terms of order \(\omega^3\), we obtain  
\bl
(\delta T_1 - \delta T_2)_{FD} &= \frac{4 \omega \mathcal{A}}{c^2} \sin(\theta + \mu) - \frac{\omega \mathcal{A}}{c^4} \frac{G M}{5 R} \big(36 \cos \theta \sin \mu - 2 (5 + 9 \sin \theta) \cos \mu \big),
\label{aboveformulaspecific}
\el
where the first term is of order \(c^{-2}\) and the second term is of order \(c^{-4}\).  
We plot these two terms separately in Fig. \ref{fig}(a) and Fig. \ref{fig}(b), respectively.
For a given polar angle \(\theta\), adjusting the Euler angle \(\mu\) will always lead to both a peak and a valley in the time delay.  To observe the effect of the second term in \eqref{aboveformulaspecific}, one can choose the orientation of the interferometer such that \(\mu + \theta = \frac{\pi}{2}\),  
which causes the first term to vanish.

\subsection{Paths $ABC$
and $ADC$: frame dragging and redshift time delays}\label{timedelayredshift}

In the last subsection, we noted that the redshift time delay vanishes for the paths \(\mathcal{CC}: ABCDA\) and \(\mathcal{C}: ADCBA\). To encode this time delay into the coincidence probability, we consider the photon paths \(\mathcal{CC}: ABC\) and \(\mathcal{C}: ADC\), while keeping the interferometer configuration unchanged, as depicted in Fig. \ref{intermoreferconfing}(b) and Fig. \ref{intermorefer}(a). In this case, both the frame dragging time delay and the redshift time delay appear.

The effective time delay due to the frame dragging effect is 
\bl
(\delta T_1 - \delta T_2)_{FD} = \frac{1}{2} \times \eqref{aboveformula}, \label{aboveformula2}
\el
which is given by the sum of \eqref{adddsa}, \eqref{adddtp}, \eqref{addgeo}, and \eqref{addlt}. The factor \(1/2\) in \eqref{aboveformula2} arises because the travel distance of each photon in this subsection is half that in Sec. \ref{frg}. Since the trajectory is not closed, Stokes’ theorem cannot be applied, thus, the derivation of \eqref{aboveformula2} is more involved, as the integration must be carried out separately along the four arms (see Appendix \ref{Appas} for details).

Using \eqref{finalequation2} and  \eqref{C13}$\sim$\eqref{tiemeDCam}, 
the difference in the redshift time delays between paths \(\mathcal{CC}\) and \(\mathcal{C}\) is 
\begin{align}
(\delta T_1 - \delta T_2)_{RS} &\equiv \frac{1}{c^3} \int_{\mathcal{CC}} d\vec{x} \cdot \hat{k} \, (\vec{\gamma} \cdot \vec{x}) 
- \frac{1}{c^3} \int_{\mathcal{C}} d\vec{x} \cdot \hat{k} \, (\vec{\gamma} \cdot \vec{x}) \notag \\
&= -\frac{2 \mathcal{A}}{c^{3}} \frac{G M}{R^{2}} (f_{2,1} - f_{1,1}) 
- \frac{2 \mathcal{A}}{c^{3}} \omega^{2} R \sin \theta \left[ \sin \theta (f_{2,1} - f_{1,1}) + \cos \theta (f_{2,2} - f_{1,2}) \right], \label{RScent}
\end{align}
where the subscript \(RS\) denotes the ``redshift". Here, \(f_{1,i}\) and \(f_{2,i}\) are the slopes of the arms \(AB\) and \(BC\), defined by \(\hat{L}_{AB} = f_{1,i} \hat{e}_{x}^{i}\) and \(\hat{L}_{BC} = f_{2,i} \hat{e}_{x}^{i}\). 
These can be read off from \eqref{uniCB} and \eqref{uniBA}, respectively.
\begin{figure}[htb]
  \centering
  \subcaptionbox{}
    {%
      \includegraphics[width = .42\linewidth]{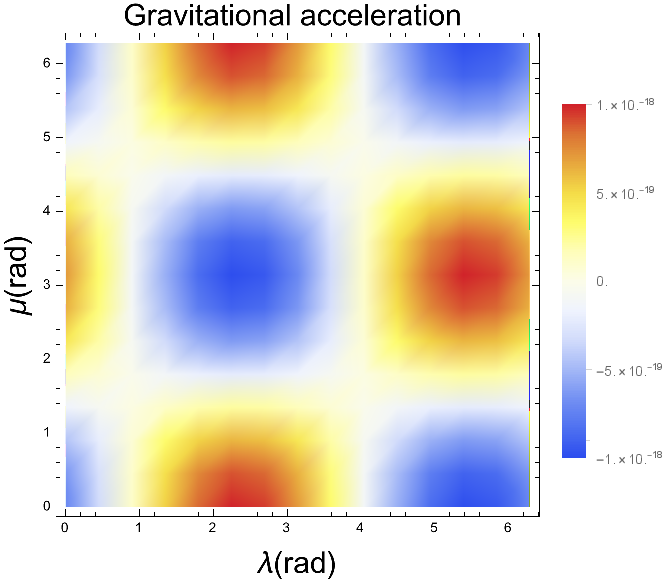}}
      \hspace{.5in}
  \subcaptionbox{}
    {%
      \includegraphics[width = .42\linewidth]{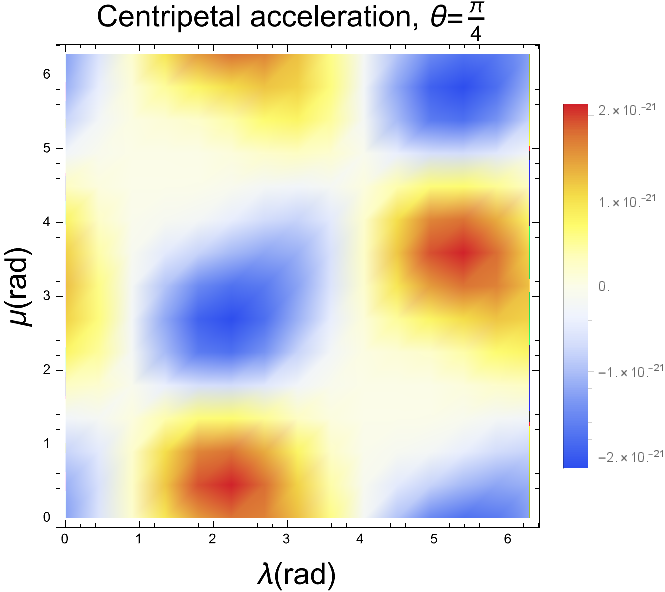}}
 \caption{ (a) Redshift time delay induced by gravitational acceleration.  
 (b) Redshift time delay induced by centripetal acceleration.
  Interferometer area: $1 \text{km}^2$.  Color: time delay, unit: s.} \label{fig2}
 \end{figure}
The first term in \eqref{RScent} arises from the Earth's gravitational potential and is independent of the polar angle \(\theta\) because the magnitude of gravitational acceleration is approximately constant on the Earth's surface. The second term depends on \(\theta\) since the centripetal acceleration of the laboratory's origin varies with the laboratory’s latitude.
Using the relevant physical parameters of the Earth provided in Appendix \ref{Appanlase}, 
the magnitudes of the time delays induced by gravitational acceleration and centripetal acceleration in \eqref{RScent}, per square kilometer, are, respectively,
\begin{equation}
10^{-18} \text{ s/km}^2, \quad \text{and} \quad 10^{-21} \text{ s/km}^2,
\label{timedelyduetoc}
\end{equation}
which are smaller than the leading order Sagnac effect, but larger than the other frame dragging contributions.

By setting the angles \(\nu=0\), \(\lambda=\alpha\), and \(\mu=\beta\) in \eqref{aboveformulaspecific}, we obtain
\bl
(\delta T_1 - \delta T_2)_{RS} &= -\frac{2 \mathcal{A}}{c^3} \frac{G M}{R^2} \cos \beta (\cos \alpha - \sin \alpha) 
- \frac{2 \mathcal{A}}{c^3} \omega^2 R \sin^2 \theta \cos \beta (\cos \alpha - \sin \alpha) \nn \\
&\quad - \frac{\mathcal{A}}{4 c^3} \omega^2 R \sin 2 \theta \, \sin 2 \beta \, (\sin 2 \alpha + \cos 2 \alpha - 1), \label{RSspecific}
\el
which is consistent with (3.6) of Ref. \cite{Brandy2021}, except for the last term. This difference arises because the second term of the centripetal acceleration in Ref. \cite{Brandy2021} is incorrect. We verified that their result can be reproduced by using the centripetal acceleration as presented in their work.

As an illustration, we plot the gravitational and centripetal acceleration-induced redshift time delays with \(\nu=0\) and \(\theta = \frac{\pi}{4}\) in Fig.\ref{fig2}(a) and Fig.\ref{fig2}(b), respectively. The gravitational acceleration-induced redshift effect is symmetric about \(\mu = \pi\). However, because the centripetal acceleration depends on the laboratory's latitude, the corresponding redshift time delay is not symmetric when \(\theta \neq 0\).

\subsection{Feasibility of experiments }\label{Amplify}

In current experiments, the accuracy of the time delay is approximately \(10^{-18}~\text{s}\) \cite{Lyons}, 
and the area of the interferometer is around \(10^{2}~\text{m}^2\) \cite{Silvestri20221, Silvestri20222, Silvestri2023}. 
The leading order Sagnac effect in \eqref{timedesagc1} and 
the gravitational acceleration-induced redshift effect in \eqref{RScent}
are the most likely to be detected. Both effects are proportional to the interferometer area, which can be effectively amplified by 
increasing the number of photon loops. To reduce the interferometer area to the currently achievable size while 
maintaining the time delay within the current precision, we shall estimate the minimum number of loops 
required for the two scenarios discussed in
Sec. \ref{frg} and Sec. \ref{timedelayredshift}.

Under the paths \(\mathcal{CC}: ABCDA\)
and \(\mathcal{C}: ADCBA\), the redshift effect  
disappears. Increasing the number of loops that photons travel through the interferometer amplifies the  Sagnac time delay
as follows
\begin{equation}
(\delta T_1 - \delta T_2)_{\text{Sag}} = n \times \left( \frac{\mathcal{A}}{10^{6}~\text{m}^2} \right) \times 10^{-15}~\text{s}, \label{AmSAg}
\end{equation}
where \(n\) is the number of loops. 
Taking \(n=10\) and \(\mathcal{A} = 10^{2}~\text{m}^2\) 
allows achieving a time precision of approximately \(10^{-18}~\text{s}\). For an equal-arm interferometer with  
\(L_{AB} = L_{BC} = 10~\text{m}\), 
a photon undergoing 10 loops should spend about \(1.3~\mu\text{s}\) in the interferometer.

Under the paths \(\mathcal{CC}: ABC\)
and \(\mathcal{C}: ADC\), the leading order Sagnac effect  is larger than the 
redshift effect. To observe the gravitational acceleration-induced redshift effect, 
according to \eqref{aboveformula}, one can orient the areal vector
\(\hat{\mathcal{A}}\) perpendicular to Earth’s angular velocity vector \(\vec{\omega}\).    
Then the remaining redshift effect can be amplified by
\begin{equation}
(\delta T_1 - \delta T_2)_{\text{RS}} \approx n \times \frac{\mathcal{A}}{10^{6}~\text{m}^2} \times 10^{-18}~\text{s}. \label{Amg}
\end{equation}
Taking \(n=10^{4}\) and \(\mathcal{A}=10^{2}~\text{m}^2\) 
makes it possible to achieve a time precision of about \(10^{-18}~\text{s}\).
For an equal-arm interferometer with \(L_{AB} = L_{BC} = 10~\text{m}\), 
to undergo \(10^{4}\) loops, a single photon should spend approximately \(0.6~\text{ms}\) in the interferometer.

\section{Detection of gravitational effects through HOM interference patterns
}\label{interference}

In this section, we revisit the HOM experiment conducted on a rotating platform, 
as described in Ref. \cite{Restuccia2019}, to determine which perspective in Sec.\ref{homdae} 
provides a correct description of the HOM experiment in the presence of the gravitational field. 
Then, we will investigate how the gravitational field of Earth influences the HOM interference patterns.

\subsection{Revisit the HOM experiment on a rotating platform}\label{revisit}

For the HOM experiment conducted on a rotating platform, see Fig. 3 in Ref. \cite{Restuccia2019}. 
Only the leading order Sagnac effect in \eqref{aboveformula} contributes to the time delay.  
When the platform is rotating clockwise, but the coupler is on the counterclockwise path, which determines the orientation of the areal vector, one has 
{\bl
\vec{\cal A}\cdot \vec{\omega} = - {\cal A}\omega,\label{atimsomeg}
\el}
where $\vec{\omega}$ is the angular velocity of the platform.
Plugging the leading order Sagnac effect of \eqref{aboveformula}
into \eqref{4666} and \eqref{68huiwh} yields 
\bl
P^{c}_{HOM}({\cal T}+\delta t_1-\delta t_2)=\frac12[1+e^{-\zeta^2[{\cal T}-4\frac{\cal A
\omega}{c^2}]^2}]
, \label{eqaiotn40HOMr}
\el
and
\bl
P^{c}_{HOM}({\cal T}+\delta T_1-\delta T_2)=\frac12[1-e^{-\zeta^2[{\cal T}+4\frac{\cal A
\omega}{c^2}]^2}]
, \label{eqaiotn40HOMl}
\el
where the coincidence probability \eqref{jointcoindeiProHOM} has been used, and
the second order time delays have been ignored.
\eqref{eqaiotn40HOMr} and \eqref{eqaiotn40HOMl} are obtained using the two-photon state constructed from the relativistic time delays and the effective time delays, respectively,
and they shift in opposite directions.

In Ref. \cite{Restuccia2019}, the authors first adjusted the coupling optics to locate 
${\cal T}$ at the steepest point on the HOM interference pattern when the platform was not rotating. They then rotated the platform clockwise and counter-clockwise, respectively, and recorded the coincidence probabilities (and the corresponding photon delays). After taking half the difference between the clockwise and counter-clockwise photon delays, the authors found that the photon delay was positive and increased with rotation speed. This experimental result can be explained using the coincidence probability \eqref{eqaiotn40HOMl}, which corresponds to equation (8) of Ref. \cite{Restuccia2019}. In fact, the interpretation of the experiment in Ref. \cite{Restuccia2019} is a coincidence, as they constructed the two-photon state using the relativistic time delay, so the coincidence probability they used is \eqref{eqaiotn40HOMr} rather than \eqref{eqaiotn40HOMl}.

The experiment in Ref. \cite{Restuccia2019} supports the view that the quantum state in curved spacetime should be constructed using the phase shift (i.e., the effective time delay) rather than the relativistic time delay. This indicates that, when studying 
a photon system within the framework of general relativity, the wave picture of photon is more adequate than the particle picture.

\subsection{HOM interference pattern in the gravitational field}

In the last subsection, we pointed out that the coincidence probability \eqref{68huiwh} can adequately describe the experimental results. In what follows, we shall show the impact on this coincidence probability caused by, for instance, the leading order Sagnac effect (the gravitational redshift effect can be treated similarly). Replace $(\delta T_1 - \delta T_2)$ by $\Delta T$ in \eqref{68huiwh} and vary ${\cal T}$. Here, $\Delta T$ is the Sagnac delay \eqref{AmSAg} or the gravitational acceleration delay \eqref{Amg}. The entire interference pattern will appear. This procedure can be applied to both spectral profiles \eqref{spectrumHOM} and \eqref{quspectrum}.

\begin{figure}[htb]
 \centering
 \subcaptionbox{}
   {%
     \includegraphics[width = .455\linewidth]{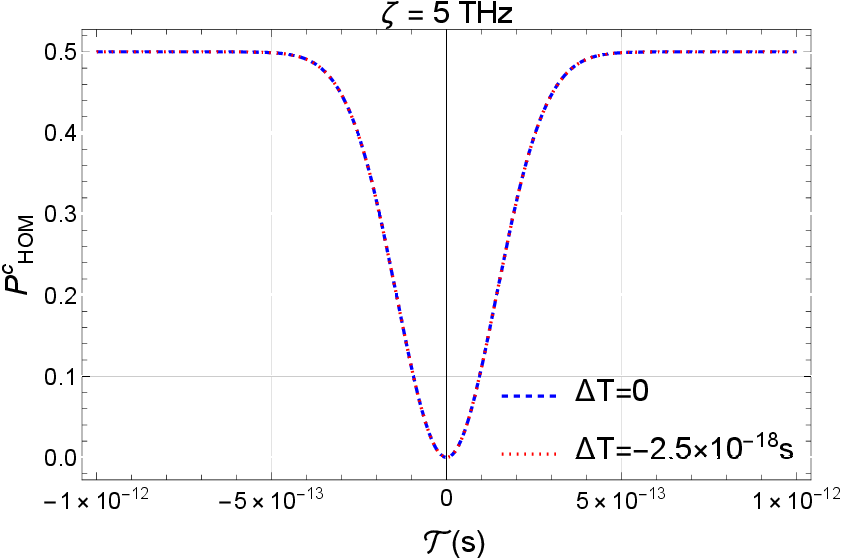}}
 \subcaptionbox{}
   {%
     \includegraphics[width = .475\linewidth]{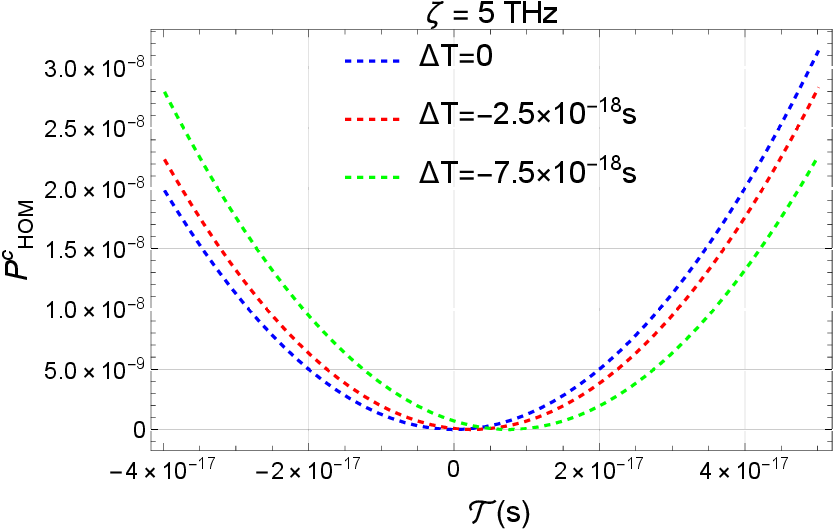}}
 \subcaptionbox{}
   {%
     \includegraphics[width = .49\linewidth]{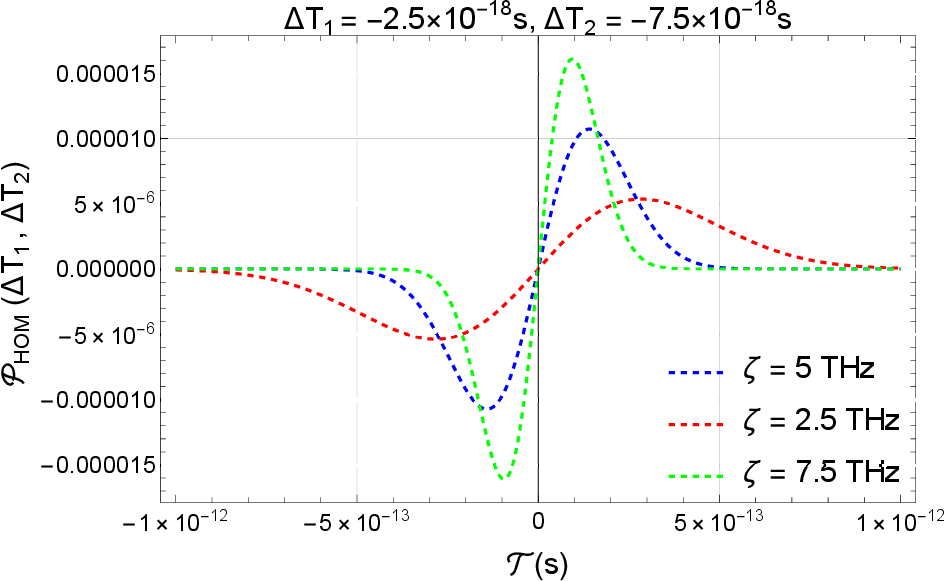}}
\caption{
(a) Entire original HOM pattern of \eqref{eq27} with and without the effective 
 time delay.
(b) Original HOM pattern of \eqref{eq27} with various effective time delays around the minimal value.
(c) ${\cal P}_{\text{HOM}}({\cal T},-2.5\times10^{-18}~\text{s},-7.5\times10^{-18}~\text{s})$ 
 with various $\zeta$.
  } \label{fig4}
\end{figure}

For the original HOM case, { replace the argument $(t_1 - t_2)$ with $({\cal T} + \Delta T)$ in the coincidence probability \eqref{jointcoindeiProHOM}, then plot the resulting function in Fig.\ref{fig4}(a) using a bandwidth of $\zeta = 5$ THz \cite{Hong1978}.}
For comparison, also plot $P^c_\text{HOM}({\cal T})$, which has no gravitational effects. Fig.~\ref{fig4}(b) is an enlargement of Fig.~\ref{fig4}(a). The two HOM patterns almost overlap because the shift $\Delta T$ is $\sim 10^{-18}\ \text{s}$ due to the smallness of the gravitational effects. Therefore, directly observing the relativistic influences through the interference pattern is difficult.

To address this, we introduce the following quantity
\bl
{\cal P}_{\text{HOM}}({\cal T},\Delta T_1,\Delta T_2) \equiv \Delta P^{c}_{\text{HOM}}({\cal T}+\Delta T_1) - \Delta P^{c}_{\text{HOM}}({\cal T}+\Delta T_2),
\label{dedifferenHOm}
\el
where $\Delta P^{c}_{\text{HOM}}$ is defined as
\bl
\Delta P^{c}_{\text{HOM}}({\cal T}+\Delta T) \equiv P^{c}_{\text{HOM}}({\cal T}+\Delta T) - P^{c}_{\text{HOM}}({\cal T}),
\label{differenHOm}
\el
and $\Delta T_1$ and $\Delta T_2$ are distinct effective time delays obtained by rotating the interferometer in different directions.

Notice that the quantity \eqref{differenHOm} is not a good probe to measure the gravitational effects in the original HOM coincidence probability, since the gravitational field of the Earth cannot be shielded. Therefore, one can never measure $P^{c}_{\text{HOM}}({\cal T})$ in a terrestrial laboratory. However, ${\cal P}_{\text{HOM}}({\cal T}, \Delta T_1, \Delta T_2)$ is a better probe to measure the gravitational effects, as $P^{c}_{\text{HOM}}({\cal T})$ is canceled in the expression \eqref{dedifferenHOm}. For illustration, we plot ${\cal P}_{\text{HOM}}({\cal T}, -2.5 \times 10^{-18}\ \text{s}, -7.5 \times 10^{-18}\ \text{s})$ with the bandwidths $\zeta = 2.5$, 5, and 7.5 THz in Fig.~\ref{fig4}(c).

Expanding \eqref{dedifferenHOm} for small $\Delta T_1$ and $\Delta T_2$, one has 
\bl
{\cal P}_{\text{HOM}}({\cal T}, \Delta T_1, \Delta T_2) \simeq {\cal T} \zeta^2 e^{-{\cal T}^2 \zeta^2} (\Delta T_1 - \Delta T_2).
\label{deltapcorigion}
\el
As ${\cal T}$ increases, the amplitude of ${\cal P}_{\text{HOM}}$ first increases, then decreases, reaching its maximal value at
\bl
|{\cal T}| = \frac{1}{\sqrt{2}\zeta},
\label{eq83}
\el
which depends only on $\zeta$. The corresponding maximal value is 
\bl
|{\cal P}_{\text{HOM}}^{\text{Max}}| \simeq \frac{1}{\sqrt{2 e}} \zeta (\Delta T_1 - \Delta T_2).
\label{maximavaluePhom}
\el
A higher peak in ${\cal P}_{\text{HOM}}({\cal T}, \Delta T_1, \Delta T_2)$ is easier to detect experimentally. To enhance this, one can either increase the difference between the two time delays $(\Delta T_1 - \Delta T_2)$ or increase the variance $\zeta$, as shown in Fig.~\ref{fig4}(c). However, as shown by \eqref{eq83}, a larger $\zeta$ requires a smaller ${\cal T}$, which in turn demands higher temporal resolution for experiments.

In practice, a scheme similar to that in Ref.~\cite{Restuccia2019} can be used to measure the change in the coincidence probability near ${\cal T} = \frac{1}{\sqrt{2}\zeta}$. This approach ensures high-precision measurements while reducing experimental workload. Detection of the peak of ${\cal P}_{\text{HOM}}({\cal T}, \Delta T_1, \Delta T_2)$ in the experiment would indicate that the effects of a slightly curved spacetime can be observed via HOM interference, and that quantum field theory in curved spacetime remains valid when considering the quantum field in the vicinity of the Earth.

\begin{figure}[htb]
  \centering
  \subcaptionbox{}
    {%
      \includegraphics[width = .46\linewidth]{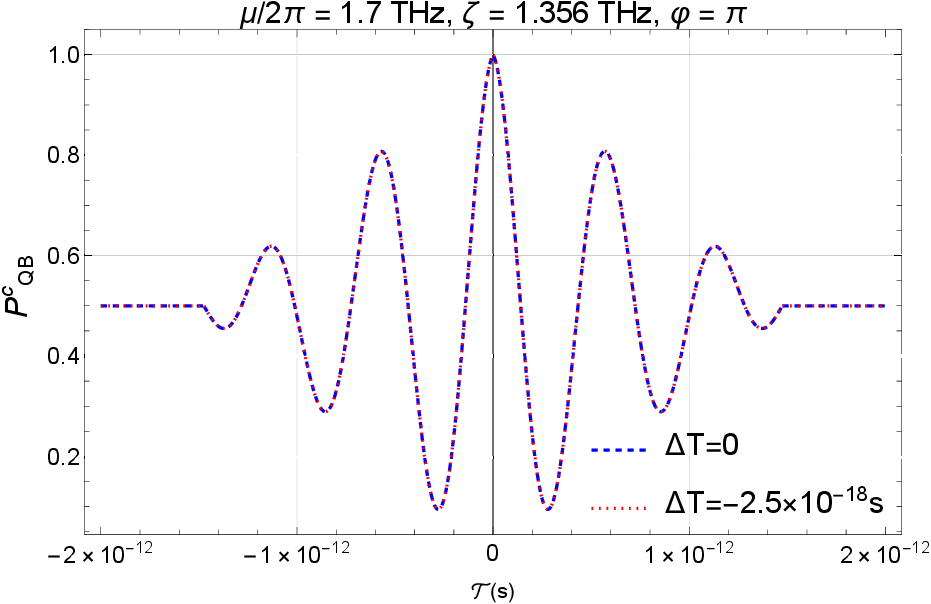}}
  \subcaptionbox{}
    {%
      \includegraphics[width = .47\linewidth]{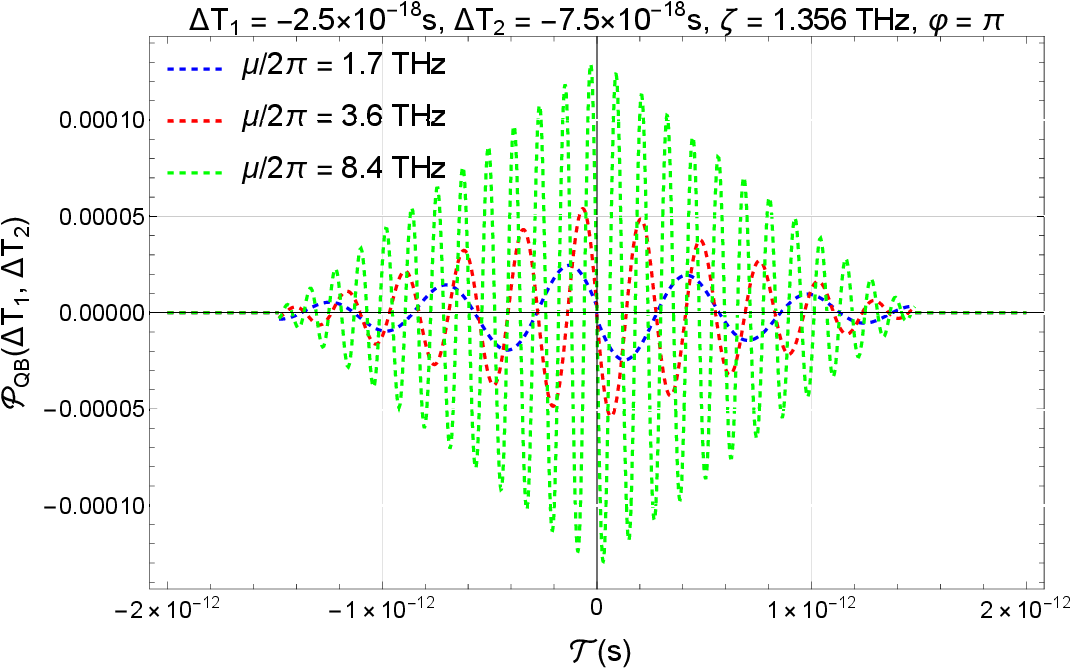}}
      \subcaptionbox{}
    {%
      \includegraphics[width = .49\linewidth]{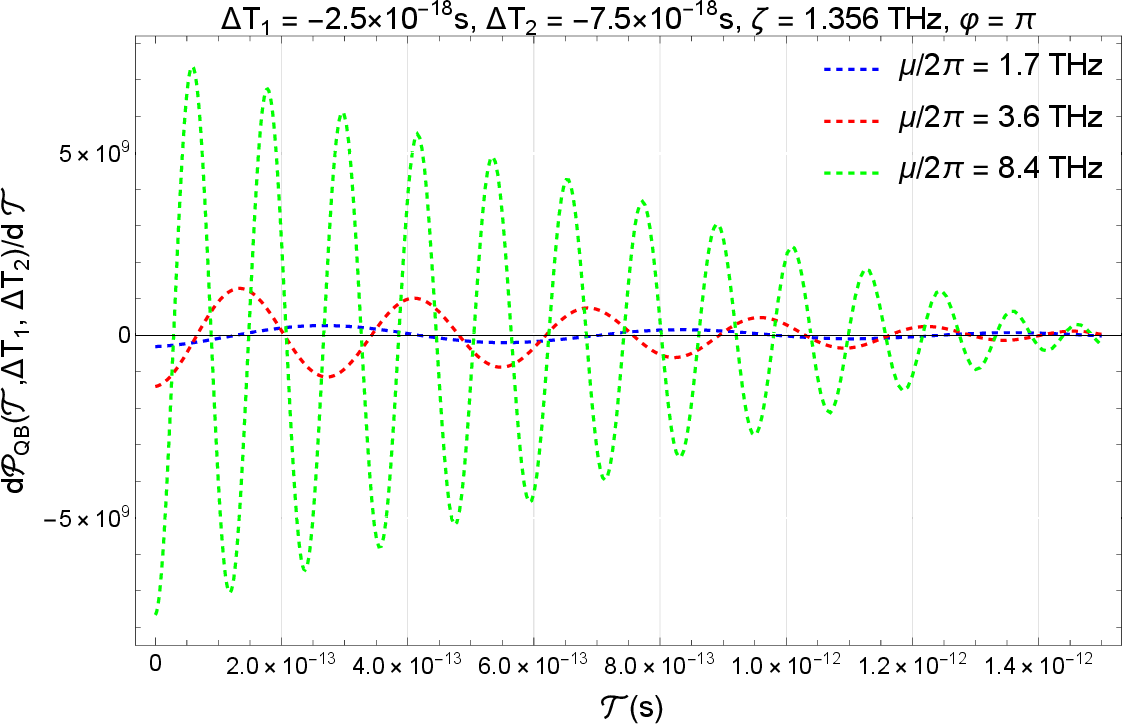}}
      \caption{
 (a) Entire quantum beating pattern of \eqref{jointcoindeiProQB} with and without
  the effective time delay.
 (b) ${\cal P}_{\text{QB}}({\cal T}, -2.5\times10^{-18}~\text{s},-7.5\times10^{-18}~\text{s})$ 
 with various $\mu$. (c). $d {\cal P}_{\text{QB}}/d{\cal T}$, which exhibits the 
 position of the peak and valley in ${\cal P}_{\text{QB}}$.
  } \label{fig5}
 \end{figure}

In the following, we investigate the quantum beating case, which involves spectrally indistinguishable and frequency-entangled photons. These photons possess additional quantum properties compared to the original HOM case. The interference pattern of the quantum beating case exhibits rapid transitions between photon bunching ($P^{c}_{QB}<0.5$) and anti-bunching ($P^{c}_{QB}>0.5$). Similar to the original HOM case, we introduce the following quantity
\bl
{\cal P}_{\text{QB}}({\cal T}, \Delta T_1,\Delta T_2)\equiv\Delta P^{c}_{QB}({\cal T}+\Delta T_1)
-\Delta P^{c}_{QB}({\cal T}+\Delta T_2), \label{dedifferenQB}
\el
where 
\bl
\Delta P^{c}_{QB}({\cal T}+\Delta T)\equiv P^{c}_{QB}({\cal T}
+\Delta T)-P^{c}_{QB}({\cal T}),\label{differenQB}
\el
and $P^{c}_{QB}$ is defined in \eqref{jointcoindeiProQB}.
In this case, the change in the coincidence probability is also difficult to measure by direct observation, as shown in Fig.\ref{fig5}(a) using parameters from Refs.~\cite{Brazel1,Ramelow2009}. 
We plot ${\cal P}_{\text{QB}}({\cal T}, \Delta T_1, \Delta T_2)$, defined in \eqref{dedifferenQB}, for various frequency separations $\mu$ in Fig.~\ref{fig5}(b). The figure shows that the oscillation frequency and the amplitude of ${\cal P}_{\text{QB}}({\cal T}, \Delta T_1, \Delta T_2)$ increase with increasing $\mu$.

Expanding \eqref{dedifferenQB} for small $\Delta T_1$ and $\Delta T_2$, one obtains
\bl
{\cal P}_{\text{QB}}({\cal T},\Delta T_1,\Delta T_2)
&\simeq\frac{
\zeta  \cos (\mu{\cal T} )-\zeta  \mu{\cal T}  \sin (\mu{\cal T})
-\zeta  \cos (\mu
(2-\zeta {\cal T})/\zeta)+2 \mu \sin (\mu{\cal T} )}
{4 (\text{sinc}(2 \mu /\zeta )-1)}(\Delta T_1-\Delta T_2),\label{1089}
\el
where ${\cal T} > 0$ and ${\cal T} + (\Delta T_1 - \Delta T_2) > 0$ have been assumed. The small-argument expansion of ${\cal P}_{\text{QB}}$ under other conditions can be obtained through a similar process and does not significantly differ from \eqref{1089}, so we omit them here.
Similar to the original HOM case, ${\cal P}_{\text{QB}}$ can be amplified by increasing $(\Delta T_1 - \Delta T_2)$, $\zeta$, and $\mu$. Simultaneously, the oscillation frequency also increases with $\mu$, as depicted in Fig.~\ref{fig5}(b). However, unlike the original HOM case, the locations of the peaks and valleys of ${\cal P}_{\text{QB}}$ cannot be determined analytically. Thus, we plot the first order derivative of \eqref{1089} with respect to ${\cal T}$ in Fig.~\ref{fig5}(c).
The intersections of $d{\cal P}_{\text{QB}}/d{\cal T}$ with the horizontal axis correspond to the positions of the peaks and valleys of ${\cal P}_{\text{QB}}$. 
{One can verify that the number of intersections of  $d{\cal P}_{\text{QB}}/d{\cal T}$
with the horizontal axis matches the total number of peaks and valleys of ${\cal P}_{\text{QB}}$.}
Similar to the approach in Ref.~\cite{Restuccia2019}, one can measure the change in the coincidence probability near the peaks or valleys, thereby ensuring high-precision measurements while reducing the experimental workload.

Ref. \cite{Brazel2} studied the HOM experiment in a satellite-terrestrial laboratory system, where they observed that the maximum value of $\Delta P^{c}_{\text{HOM}}$ is independent of $\zeta$. Thus, Ref. \cite{Brazel2} claimed that adjusting $\zeta$ cannot amplify $\Delta P^{c}_{\text{HOM}}$.
In the quantum beating case with the spectrum \eqref{quspectrum}, Ref.~\cite{Brazel2} suggested that $\Delta P^{c}_{\text{QB}}$ is a more efficient probe for detecting gravitational effects than $\Delta P^{c}_{\text{HOM}}$, as the maximum value of $\Delta P^{c}_{\text{QB}} \sim \frac{1}{\zeta}$ can be enhanced by reducing $\zeta$. However, our work shows that the maximum value of $\Delta P^{c}_{\text{HOM}}$ can be enhanced by increasing $\zeta$ to a level comparable to that of $\Delta P^{c}_{\text{QB}}$. Therefore, the original HOM and the quantum beating cases are both suitable for studying gravitational effects in HOM experiments, which is an advantage of using the interferometer depicted in Sec.~\ref{expsetup}. In Ref.~\cite{Brazel2}, the gravitational effect is reflected by the redshift $z$ in the term $\zeta^{2}(1+z)^{2}\Delta \tau^{2}$, instead of $\zeta^{2}({\cal T} + \Delta T)^{2}$ as used in this paper, where $\Delta \tau$ is an alternative time delay equivalent to ${\cal T}$. The maximum value of $\Delta P^{c}_{\text{HOM}}$ is independent of $\zeta$ in Ref.~\cite{Brazel2} because it occurs at $\Delta \tau_{\max} \sim 1/\zeta$, which cancels the $\zeta^{2}$ factor in $\zeta^{2}(1+z)^{2} \Delta \tau^{2}$. However, in our case, since ${\cal T} \sim \frac{1}{\zeta}$ (see \eqref{eq83}), it does not get canceled in $\zeta^{2}({\cal T} + \Delta T)^{2}$. This is why the maximum value of $\Delta P^{c}_{\text{HOM}}$ in \eqref{differenHOm} depends on $\zeta$.

Detecting the leading order Sagnac and gravitational redshift effects through the coincidence probability $P^{c}$ is challenging since it requires a high temporal resolution on the order of $\sim 10^{-18}~\text{s}$. However, a notable advantage of measuring the quantities ${\cal P}_{\text{HOM}}$ and ${\cal P}_{\text{QB}}$ is that they require only a temporal resolution on the order of $\sim 10^{-12}~\text{s}$. Nevertheless, measuring these quantities places higher demands on single-photon generation and photon counting technologies.

\section{Conclusions and discussion}\label{conclusionanddis}

In this paper, we investigate how Earth's gravitational field affects the HOM interference experiments conducted in a terrestrial laboratory. We calculate the time delays up to second order from two perspectives. For the particle perspective, where the photon is localized, the relativistic time delays \eqref{0thapp21} and \eqref{dede1} are obtained by solving the null geodesic equation \eqref{lightinterval} for a massless particle. Nevertheless, for the wave perspective, where the photon is non-localized, the effective time delays \eqref{phasetotime11} and \eqref{phasetotime22} can be derived using the phase shifts \eqref{S1} and \eqref{S2}, which are obtained by solving the massless minimally-coupling Klein-Gordon equation. We find that the relativistic time delay differs from the effective time delay by several terms at the second order, and they differ by a minus sign even at first order, because both the temporal and spatial parts of the phase shift are influenced by gravity.
We point out in Appendix \ref{Appcontoti} that the relativistic time delay cannot be trivially obtained by dividing $(c k_0)$ by the phase shift, they are actually related by a complicated coefficient vector. By revisiting the HOM experiment conducted on a rotating platform, we find that the experimental results support the wave perspective. In this regard, 
the wave picture is more adequate  than the particle picture.

From the wave perspective, the first order effective time delay \eqref{S1} consists of the frame dragging part arising from the $g_{0i}$ component and the redshift part arising from the $g_{00}$ component, as classified in Ref.\cite{Brandy2021}. The frame dragging time delay results from the leading and next-to-leading order Sagnac effects, the Geodetic effect, the Lense-Thirring effect, and the Thomas precession effect. The next-to-leading order Sagnac effect, which results from the coupling of rotation and acceleration, had been previously overlooked in Refs.\cite{Brandy2021, Bosi2009}. However, for consistency, this effect should be taken into account, as its order of magnitude is comparable to that of the other effects except for the leading order Sagnac effect. The redshift time delay is contributed by both the gravitational and the centripetal accelerations.
We study the frame dragging and redshift effects within an arbitrarily oriented closed rectangular interferometer in two distinct scenarios. In the first case, the light paths are ${\cal CC}:ABCDA$ and ${\cal CC}:ADCBA$, where the frame dragging effect is present and the redshift effect vanishes. In the second case, the light paths ${\cal CC}:ABC$ and ${\cal CC}:ADC$ are chosen, where both the frame dragging and redshift effects occur, with the frame dragging effect being half of that in the first case. To isolate the redshift effect, one can set $\hat{\cal A} \cdot \vec{\omega} = 0$ to ensure that the frame dragging effect is zero. These two scenarios allow the frame dragging and redshift effects to be studied separately.

The order of magnitude of the leading order Sagnac effect is approximately $10^{-15}~\text{s/km}^2$, 
and the redshift effect caused by gravitational acceleration is about $10^{-18}~\text{s/km}^2$, 
both of which are potentially detectable. 
To achieve the current time precision of $\sim 10^{-18}$ s \cite{Lyons} within a proper interferometer size of $\sim 10^2$ m$^2$ \cite{Silvestri20221, Silvestri20222}, 
the number of light loops should be 10 for the leading order Sagnac effect and $10^3$ for the redshift effect. As discussed in Sec.~\ref{interference}, detecting gravitational effects by directly observing the interference pattern is challenging, 
because a time delay on the order of $10^{-18}$ s causes only a slight shift in the pattern. 
To overcome this difficulty, we propose using ${\cal P}({\cal T}, \Delta T_1, \Delta T_2)$ as the observable quantity. For the original HOM case, ${\cal P}_{HOM}({\cal T}, \Delta T_1, \Delta T_2)$ exhibits a peak and a valley around ${\cal T} \sim \pm 1/(\sqrt{2}\zeta) \sim 10^{-12}$ s, 
which may be observed experimentally. The amplitude of ${\cal P}_{HOM}$ can be enhanced by increasing $(\Delta T_1 - \Delta T_2)$ and $\zeta$, 
a behavior that differs from Ref.~\cite{Brazel2} due to the different interferometer configuration used there. For the quantum beating case, ${\cal P}_{QB}$ oscillates due to the rapid transitions between the photon bunching and anti-bunching processes, 
and both the oscillation frequency and amplitude can be amplified by increasing $(\Delta T_1 - \Delta T_2)$, $\zeta$, and $\mu$. Hence, ${\cal P}_{QB}$ provides an alternative probe for measuring gravitational effects in HOM experiments. For both the original HOM and quantum beating cases, 
detecting ${\cal P}$ near its extreme points enables high-precision measurements 
while reducing experimental workload.

In the current study, the redshift and frame dragging effects due to Earth’s gravitational field are encoded in the effective time delay derived from the phase shift, influencing the HOM interference pattern. HOM interference is a distinct quantum phenomenon without a classical analogue, thus providing a valuable tool to assess gravitational effects on quantum systems. The observation of peaks or valleys in Fig.~\ref{fig4}(d), or signals in Figs.~\ref{fig5}(c) and (f), would support the validity of the formulas presented in Sec.~\ref{appraoch2} for describing the interplay between general relativity and quantum theory at low energies. These formulas could serve as a foundation for developing a more comprehensive theory. On the other hand, the absence of such signals in these figures may indicate that combining general relativity and quantum theory in this manner is not feasible, suggesting a need for theoretical modification. Regardless of the experimental outcomes, the results of the HOM interference experiments will provide new insights into the unification of general relativity and quantum theory.

\

\
\textbf{Acknowledgements}

\

We thank HuiNan Wu, Boyu Fu and Zhixuan Geng for their helpful discussion on the experiment's feasibility, and Yifu Cai for his support. This work was supported in part by the National Key R$\&$D Program of China (2024YFC2207500, 2021YFC2203100), the CAS Young Interdisciplinary Innovation Team (JCTD-2022-20), the NSFC (12261131497, 92476203), the 111 Project for “Observational and Theoretical Research on Dark Matter and Dark Energy” (B23042), the Fundamental Research Funds for Central Universities, the CSC Innovation Talent Funds, the USTC Fellowship for International Cooperation, and the USTC Research Funds of the Double First-Class Initiative.

\appendix
\numberwithin{equation}{section}

\section{Approximate value of physical quantities
}\label{Appanlase}

In this appendix, we provide the approximate values of the physical quantities,
\bl
\omega&\approx 7.3\times 10^{-5} \text{ rad/s},
\\
R&\approx 6.4\times 10^6 ~\text{m},
\\
M&\approx 6\times 10^{24} ~\text{kg},
\\
G&\approx 6.7\times 10^{-11} \text{ m}^3/(\text{kg $\cdot$ s}^{2}),
\\
c&\approx 3\times 10^8 ~\text{m}/\text{s},
\\
\frac{G M}{c^2 R}&\approx 7\times 10^{-10},
\el
where $\omega$ is the angular velocity of the Earth's rotation, $R$ is the radius of the Earth, $M$ is the mass of the Earth, $G$ is the gravitational constant, and 
$c$ is the speed of light in vacuum.

\section{Vectors expanded by different sets of  basis vectors}\label{Appcoordinate}

In this appendix, we show the relations between the basis vectors of the geocentric reference frame and those of the laboratory reference frame. Moreover, the vectors $\vec{\omega}, \vec{v}, \vec{a}_c, \nabla U, \nabla \times \vec{V}$ will be expanded in terms of these two sets of basis vectors.

Three orthonormal basis vectors of the geocentric reference frame are denoted by $\hat{e}_X^{i}$, which satisfy
\bl
\hat{e}_X^{i}\cdot \hat{e}_X^{j} = \delta^{ij}, \quad i=1,2,3,
\el
where $(\hat{e}_X^{1}, \hat{e}_X^{2}, \hat{e}_X^{3})$ coincide with the standard basis vectors $(\hat{i}, \hat{j}, \hat{k})$ of the Cartesian coordinate system in the geocentric reference frame. The Earth's equatorial plane is spanned by $\hat{e}_X^{1}$ and $\hat{e}_X^{2}$, and the Earth's angular velocity vector is $\vec{\omega} = \omega \hat{e}_X^{3}$.

Similarly, three orthonormal basis vectors of the laboratory reference frame are denoted by $\hat{e}_x^{i}$, satisfying
\bl
\hat{e}_x^{i}\cdot \hat{e}_x^{j} = \delta^{ij}, \quad i=1,2,3,
\el
where $(\hat{e}_x^{1}, \hat{e}_x^{2}, \hat{e}_x^{3})$ coincide with the standard basis vectors $(\hat{e}_r, \hat{e}_\theta, \hat{e}_\phi)$ of the spherical coordinate system in the geocentric reference frame.

{The transformations between the basis vectors $\{\hat{e}_X^i\}$ and $\{\hat{e}_x^i\}$ are given by}
\begin{subequations}
\bl
\hat{e}_X^1 &= (\sin\theta \cos\phi)\, \hat{e}_x^1 + (\cos\theta \cos\phi)\, \hat{e}_x^2 - (\sin\phi)\, \hat{e}_x^3, \label{inverrelation1123}
\\
\hat{e}_X^2 &= (\sin\theta \sin\phi)\, \hat{e}_x^1 + (\cos\theta \sin\phi)\, \hat{e}_x^2 + (\cos\phi)\, \hat{e}_x^3, 
\label{inverrelation2123}
\\
\hat{e}_X^3 &= (\cos\theta) \hat{e}_x^1 - (\sin\theta) \hat{e}_x^2.\label{inverrelation3123}
\el
\end{subequations}
The coordinate transformation between the geocentric and laboratory coordinate systems is \cite{FuKashima}
\begin{subequations}
\bl
\vec{X} &= \vec{R} + \vec{x} + O(c^{-2}), 
\\
X^0 &= x^0 + O(c^{-2}),
\el
\end{subequations}
where $\vec{X}$ is a vector in the geocentric coordinate system, and $\vec{R}$ denotes the vector from the center of the Earth to the origin of the interferometer, with magnitude $|\vec{R}|$ equal to the Earth's radius. The vector $\vec{x}$ corresponds to the vector in the laboratory coordinate system. The coordinate times in the geocentric and laboratory frames are $X^0/c$ and $x^0/c$, respectively.

The vectors $\vec{R}$ and $\vec{x}$ are expanded in terms of their respective basis vectors as
\begin{subequations}
\bl
\vec{R} &= R \hat{N}_i \hat{e}_X^i, \label{Riintermodorighn} 
\\
\vec{x} &= x^i \hat{e}_x^i,\label{xintermodorighn}
\el
\end{subequations}
where $\hat{N}_i \equiv (\sin\theta \cos\phi, \sin\theta \sin\phi, \cos\theta)$, and $x^i$ are the spatial coordinates of a point in the laboratory coordinate system.

Expressing the vectors $\vec{\omega}, \vec{v}, \vec{a}_c, \nabla U,$ and $\nabla \times \vec{V}$ in terms of the basis vectors $\{\hat{e}_X^i\}$, we have
\begin{subequations}\bl
\vec{\omega}&=\omega \hat{e}_X^3,\label{intermsofgeo}
\\
\vec{v}&=\vec{\omega}\times \vec{R}=\omega R (\hat{N}_1\hat{e}_X^2-
\hat{N}_2\hat{e}_X^1),\label{intermsofv}
\\
\vec{a}_c&=c^2\frac{d^2\vec{R}}{d(X^0)^2}=-\omega^2R(\hat{N}_1\hat{e}_{X}^{1}
+\hat{N}_2\hat{e}_{X}^{2}),\label{intermsofacc}
\\
\nabla U&=\frac{GM }{R^2}\hat{N}_i\hat{e}_{X}^i,\label{intermsofgrav}
\\
\nabla \times \vec{V}&=-\frac12\frac{GI\omega}{R^3}\Big[ -3 \hat{N}_1 \hat{N}_3\hat{e}_{X}^1-3\hat{N}_2 \hat{N}_3\hat{e}_{X}^2
+(1-3\hat{N}_3^2)\hat{e}_{X}^3\Big],\label{intermsofavec}
\el
\end{subequations}
where $\omega$ is assumed constant in time.

Plugging \eqref{inverrelation1123}, \eqref{inverrelation2123}, and \eqref{inverrelation3123}
into \eqref{intermsofgeo}, \eqref{intermsofv}, \eqref{intermsofacc}, \eqref{intermsofgrav}, and \eqref{intermsofavec}, 
the vectors can
be expressed in terms of the basis vectors  $\{\hat{e}_x^i\}$ as
\begin{subequations}
\bl
\vec{\omega}&=\omega (\cos\theta \hat{e}_x^1
-\sin\theta\hat{e}_x^2),\label{intermsofgeoab}
\\
\vec{v}&=\omega R \sin\theta e_x^3,\label{intermsofvab}
\\
\vec{a}_c&=-\omega^2R\sin\theta(\sin\theta \hat{e}_{x}^{1}
+\cos\theta\hat{e}_{x}^{2}),\label{intermsofaccab}
\\
\nabla U&=\frac{GM }{R^2}\hat{e}_{x}^1,\label{intermsofgravab}
\\
\nabla \times \vec{V}&=\frac12\frac{GI\omega}{R^3}\Big[ 2\cos\theta \hat{e}_{x}^1+\sin\theta \hat{e}_{x}^2\Big].\label{intermsofavecab}
\el
\end{subequations}
The centripetal acceleration given    
in Ref.\cite{Brandy2021}, $\omega^2 R 
\sin \theta\left(\sin \theta \hat{e}_z+\cos \theta \hat{e}_y\right)$, differs from \eqref{intermsofaccab}.

\section{Relate  the  metrics \eqref{observrefraenfram12} and \eqref{PNmetric}}\label{APPad}

In this appendix, we outline the derivation of relations \eqref{accerlation} and \eqref{rotation}.  
First, we calculate the tetrad associated with the observer at an initial time.  
Second, we perform a spatial rotation on the initial tetrad to obtain the tetrad at a subsequent time.  
Third, we express $(\vec{\gamma},~\vec{\omega}')$ in terms of $(U, \vec{V})$ using the tetrad at a given moment.

The initial tetrad \( e_{~\hat{\mu}}^{\alpha} \) satisfies the following relation \cite{MTW,Delva2017}
\bl
\eta_{\hat{\mu}\hat{\nu}} = e_{~\hat{\mu}}^{\alpha} e_{~\hat{\nu}}^{\beta} \mathfrak{g}_{\alpha\beta},
\label{intialrelation}
\el
where a hat $~\hat{}~$ denotes quantities related to the initial tetrad, and \(\mathfrak{g}_{\mu\nu}\) is given by \eqref{PNmetric}. Indices with hats are lowered or raised using \(\eta_{\hat{\mu}\hat{\nu}}\) or \(\eta^{\hat{\mu}\hat{\nu}}\), while indices without hats are lowered or raised using \(\mathfrak{g}_{\mu\nu}\) or \(\mathfrak{g}^{\mu\nu}\).

By setting $e^{\alpha}_{~\hat{0}}=\frac1{c}\frac{d X^{\alpha}}{d \tau}$, 
where $d\tau=c^{-1}\sqrt{-ds^2}$, and solving 
\eqref{intialrelation}, one obtains
\begin{subequations}
\bl
e^{0}_{~\hat{0}}
&=1+\frac{1}{c^2}(U+\frac{v^2}{2})
+\frac{1}{c^4}[-4V_i v^i+Uv^2 +\frac{3}{2}(U+\frac{v^2}{2})^2-U^2],\label{e00finalll}
\\
e^{i}_{~\hat{0}}&=\frac{v^i}{c}e^{0}_{~\hat{0}},\label{finalei0000}
\\
e^{0}_{~\hat{j}}&= -\frac{4V_j}{c^3}+\frac{ v_j}{c}+\frac{1}{c^3}(3U+\frac12v^2)v_j,\label{finaltratre0i}
\\
e^{k}_{~\hat{j}}&=\delta^{k}_{~j}+\frac{1}{c^2}(\frac{1}{2}v_kv_j-U\delta_{kj}),\label{eaffff}
\el
\end{subequations}
where \( v^i \) and \( U \) are the velocity and the gravitational potential of the laboratory origin in the geocentric reference frame, respectively. \eqref{e00finalll}, \eqref{finalei0000}, \eqref{finaltratre0i}, and \eqref{eaffff} are consistent with (35) in Ref.~\cite{Delva2017}. The choices of the spacelike vectors \( e^{\mu}_{~\hat{j}} \) are not unique, different choices are related by a Lorentz transformation. The leading order spatial vectors \( e^{k}_{~\hat{j}} \) in \eqref{eaffff} are parallel to the basis vectors \( \hat{e}_{X}^i \) defined in Appendix \ref{Appcoordinate}.

The relation between the initial tetrad and 
the tetrad at a later time is determined  using the spatial rotation matrix $\Lambda^{\hat{j}}_{~(i)}$, 
which satisfies the rotation equation $
 d \Lambda^{\hat{i}}_{~(j)}/d[c^{-1}X^{0}]
=-\Lambda^{\hat{m}}_{~(j)}\epsilon^{\hat{i}}_{~\hat{m}\hat{l}}\omega^{\hat{l}}$ \cite{Delva2017}. Thus, 
the tetrad at a given time $c^{-1}X^{0}$ is 
\begin{subequations}
\bl
e^{\mu}_{~(0)}&=e^{\mu}_{~\hat{0}},\label{44emujnewu}
\\
e^{\mu}_{~(i)}&=\Lambda^{\hat{j}}_{~(i)}[c^{-1}X^{0}] e^{\mu}_{~\hat{j}},\label{44emujnewd}
\el
\end{subequations}
where $()$ denotes the indices of the tetrad at time $c^{-1} X^{0}$. The indices  with $()$ are lowered or raised using $\eta_{(\mu)(\nu)}$ or $\eta^{(\mu)(\nu)}$.  
In the rest of this appendix, vectors with $()$ are expanded in the basis vectors $\hat{e}_{x}^i$.
Alternatively, the tetrad $e^{\mu}_{~(\alpha)}$ can be obtained by directly solving the equation  
$\eta_{(\mu)(\nu)} = e^{\alpha}_{~(\mu)} e^{\beta}_{~(\nu)} \mathfrak{g}_{\alpha\beta}$.
However, since the form of $e^{k}_{~(j)}$ is non-trivial and difficult to determine, we do not use this approach.


In what follows, we relate the quantities  
$(\vec{\gamma}, \vec{\omega}')$ in \eqref{observrefraenfram12} to  
$(U, \vec{V})$ in \eqref{44emujnewu} and \eqref{44emujnewd}.  
The tetrad varies from point to point along the observer’s worldline,  
relative to parallel transport \cite{MTW}
\bl
\frac{De^{\mu}_{~(\alpha)}}{D\tau}=-\Omega^{\mu\nu} e_{\nu(\alpha)},\label{tetradpaprell}
\el
where $D/D\tau$ denotes the covariant derivative along the observer’s worldline, and 
$\Omega^{\mu\nu}$ is an antisymmetric tensor defined by  
\bl \label{defianOmgea}
\Omega^{\mu\nu} = a^{\mu} u^{\nu} - u^{\mu} a^{\nu} + u_{\alpha} \omega_{\beta} \epsilon^{\alpha \beta \mu \nu},
\el
where $u^{\mu}$ and $a^{\mu} = D u^{\mu} / D \tau$ are the 4-velocity and 4-acceleration of the observer, and 
$\omega^{\mu}$ is the 4-angular velocity of $e^{\mu}_{~(j)}$ relative to Fermi-Walker-transported vectors. Both $a^{\mu}$ and $\omega^{\mu}$ are orthogonal to the 4-velocity $u^{\mu}$, i.e.,
$a^\mu u_{\mu} = 0, ~ \omega^{\mu}u_\mu = 0$.

Multiplying $e_{\mu(\theta)}e_{\nu(\gamma)}$ on both sides of $(\Omega^{\mu\nu}-\Omega^{\nu\mu})$,
and using \eqref{tetradpaprell} together with $\frac{D}{D\tau}(e_{\nu}^{~(\sigma)}e^{\nu}_{(\alpha)})=0$, one has
\bl
\Omega_{(\theta)(\gamma)}
&=\frac12\mathfrak{g}_{\mu\nu}\Big(e^{\mu}_{~(\gamma)}\frac{De^{\nu}_{~(\theta)}}{D\tau}
-e^{\mu}_{~(\theta)}\frac{De^{\nu}_{~(\gamma)}}{D\tau}\Big),\label{Omegasigama}
\el
which differs from that in Ref. \cite{Delva2017} by a negative sign due to the different metric convention used.
Using \eqref{defianOmgea} and \eqref{Omegasigama}, the acceleration $a_{(\theta)}$, as measured by the accelerometers
of the observer, is given by 
\bl
a_{(i)}&=-\Omega_{(i)(0)},\label{subscriptai}
\el
where relations $a_{(\theta)}=e_{~(\theta)}^{\mu}a_{\mu}$,  
$\eta_{(0)(0)}=-1$, $a_{(0)}=e_{\nu(0)}
\frac{D}{D\tau}e^{\nu}_{~(0)}=0$,
$\epsilon_{(0)(\beta)(\theta)(0)}=0$ have been used.
Furthermore, using \eqref{defianOmgea} and \eqref{Omegasigama}, 
the angular velocity $\Omega^{(m)}$, as measured by the gyroscopes
of the observer, is given by  
\bl
\Omega^{(m)}&\equiv \frac12\epsilon^{(i)(j)(m)}\Omega_{(i)(j)},\label{omegameasurebyobsebat}
\el
where relations $u_{(i)}=e_{\alpha(i)}e^{\alpha}_{~(0)}=0, u^{(0)}=1$ and $\epsilon_{(0)(k)(i)(j)}=\epsilon_{(k)
(i)(j)}$ have been used. By plugging  \eqref{e00finalll}$\sim$\eqref{44emujnewd} into \eqref{subscriptai}, one obtains 
\bl
a^{(m)}&=a_c^{(m)}-U_{,(m)},\label{anhatfinalnew}
\el
where $a_{c}^{(m)}$ and $U_{,(m)}$
are the centripetal and gravitational accelerations  at the laboratory origin in the laboratory reference frame, respectively, and both vectors are
expanded by using the basis vectors $\hat{e}_{x}^i$.
Renaming $a^{(m)}$ as $\gamma^m$
and expressing it in vector form, one obtains 
\eqref{accerlation}.

Plugging \eqref{44emujnewu} and \eqref{44emujnewd} into \eqref{omegameasurebyobsebat} and using the rotation equation, 
one finds, after a lengthy calculation, that
\bl
\Omega^{(m)}
&=\frac{1}{2}\Big\{
\omega^{(m)}\nn
+\frac{1}{c^2}(\frac{v^2}2+U)2
\omega^{(m)}-\frac{1}{2c^2} 
\epsilon^{(j)(i)(m)}\Big[
  v_{(j)}(U_{,(i)}-\dot{v}_{(i)})-v_{(i)}(U_{,(j)}- v_{(i)}\dot{v}_{(j)}) 
\Big]\nn
\\
&~~~~~~~~~~~~~~~~-\frac{3}{2c^2} \epsilon^{(j)(i)(m)}\Big[v_{(j)}
U_{,(i)}-v_{(i)}U_{,(j)}\Big]-\frac{2}{c^2}\epsilon^{(j)(i)(m)}
\Big[V_{(i),(j)}-V_{(j),(i)}\Big]\Big\},\label{omegineetre}
\el
where $\omega^{(m)}$, $V_{(i)}$, and $v^{(i)}$
denote the components of the angular velocity, the gravitomagnetic vector potential of the Earth, and the velocity of the laboratory origin relative to the laboratory reference frame, respectively.
Renaming $\Omega^{(i)}$ as $\omega'^{i}$ 
and expressing it in vector form leads to
\eqref{rotation}.

\section{Relations between the relativistic time delays and the phase shifts}\label{Appcontoti}

In this Appendix, we relate the relativistic time delay to the phase shift.  
Raising the index $\alpha$ of $k_{\alpha}$ as well as  \eqref{sohdieuwfh1} and \eqref{sohdieuwfh2} by using
$g^{\mu\nu} = \eta^{\mu\nu} - h^{\mu\nu} + h^{\mu\alpha} h_{\alpha}^{~\nu}$, we obtain
\begin{subequations}
\bl
[\nabla^{\mu} S]^{0}&=k^{\mu},\label{eqre93}
\\
[\nabla^{\mu} S]^{1}&=
-\frac12 h^{\mu\nu} k_{\nu},\label{eqre94}
\\
[\nabla^{\mu} S]^{2}&=\frac38 h^{\mu}_{~\alpha}h^{\alpha}_{~\rho} 
k^{\rho},\label{eqre95}
\el
\end{subequations}
where the superscript outside the square brackets denotes the perturbation order of the metric $h_{\mu\nu}$.  

Comparing \eqref{lightinterval} with \eqref{eqfjrieo}, one finds that the zeroth, first, and second order times should be expressed as combinations of phases of various orders
\begin{equation}\label{equ96fd}
\begin{aligned}
{}[d x^{0}]^{0} + [d x^{0}]^{1} + [d x^{0}]^{2} 
= \frac{c}{\omega} \big\{ [\Lambda^{0}_{~\alpha} ]^{0} + [\Lambda^{0}_{~\alpha} ]^{1} + [\Lambda^{0}_{~\alpha} ]^{2} \big\} \cdot \big\{ [\nabla^{\alpha} S]^{0} + [\nabla^{\alpha} S]^{1} + [\nabla^{\alpha} S]^{2} \big\} \, dl,
\end{aligned}
\end{equation}
where all quantities are expanded in terms of the metric perturbation.  
The zeroth, first, and second order coefficient vectors are defined as $[\Lambda^{0}_{~\alpha} ]^{0}$, $[\Lambda^{0}_{~\alpha} ]^{1}$, and $[\Lambda^{0}_{~\alpha} ]^{2}$, respectively.  
We determine them by comparing $[d x^{0}]^{0}$, $[d x^{0}]^{1}$, and $[d x^{0}]^{2}$ with equations \eqref{0thordert1}, \eqref{1stordert1}, and \eqref{2ndordert1}, respectively, as follows
\begin{subequations}\bl
[\Lambda^{0}_{~\alpha} ]^{0}&=\delta^{0}_{~\alpha}
+\frac{c}{\omega}k_{\alpha},\label{lambda0thd}
\\
[\Lambda^{0}_{~\alpha} ]^{1}&=\frac12 h^{0}_{~\alpha}+\frac{c^2}{\omega^2}
 h_{\alpha\beta} k^{\beta}k^{0},\label{lambda1std}
\\
[\Lambda^{0}_{~\alpha} ]^{2}&=\frac{1}{2}\frac{c^2}{\omega^2}h^{0\beta}h^{\rho}_{~\alpha}
k_{\beta}k_{\rho}+\frac18\frac{c^2}{\omega^2} h^{\beta\rho}h_{\rho\alpha}
k^0k_{\beta}-\frac18\frac{c^4}{\omega^4} h_{\beta\rho}h_{\sigma\alpha} 
k^0 k^{\beta}k^{\rho}k^{\sigma}
-\frac{3}{8}h^{0}_{~\sigma}h^{\sigma}_{~\alpha}.\label{lambda2ndd}
\el
\end{subequations}
Plugging \eqref{eqre93}, \eqref{eqre94}, \eqref{eqre95} and 
\eqref{lambda0thd}, \eqref{lambda1std}, \eqref{lambda2ndd} into the right hand side of \eqref{equ96fd}, one can verify that the following expressions
\begin{subequations}
\begin{align}
d\bar{t} &= \frac{1}{c}[dx^0]^{0} = \frac{1}{\omega} [\Lambda^{0}_{~\alpha}]^{0} [\nabla^{\alpha} S]^{0} dl, \label{rel1n} \\
dt^{(1)} &= \frac{1}{c}[dx^0]^{1} = \frac{1}{\omega} \Big( [\Lambda^{0}_{~\alpha}]^{0} [\nabla^{\alpha} S]^{1} + [\Lambda^{0}_{~\alpha}]^{1} [\nabla^{\alpha} S]^{0} \Big) dl, \label{rel2n} \\
dt^{(2)} &= \frac{1}{c}[dx^0]^{2} = \frac{1}{\omega} \Big( [\Lambda^{0}_{~\alpha}]^{0} [\nabla^{\alpha} S]^{2} + [\Lambda^{0}_{~\alpha}]^{2} [\nabla^{\alpha} S]^{0} + [\Lambda^{0}_{~\alpha}]^{1} [\nabla^{\alpha} S]^{1} \Big) dl, \label{rel3n}
\end{align}
\end{subequations}
coincide with the time delays given in \eqref{0thordert1}, \eqref{1stordert1}, and \eqref{2ndordert1}, respectively.
This result confirms that the times can be calculated from a nontrivial combination of the phases and cannot be simply obtained by dividing the phase by $(c k_{0})$.

\section{ Effective time delays of various light paths}\label{teimdelayvar}

In this Appendix, we calculate the effective time delays for various light paths.  
To compare with the results in Refs. \cite{Brandy2021,Bosi2009}, one should add a minus sign to the corresponding formulas  in this Appendix.

\subsection{Paths ${\cal CC}: ABCDA$
and ${\cal C}: ADCBA$}\label{c11111}

Along the paths ${\cal CC}: ABCDA$ and ${\cal C}: ADCBA$, 
only the frame dragging effect, the first term in \eqref{phasetotime11}, contributes to the time delay, which is
\bl
(\delta T_1-\delta T_2)_{FD}
&= -\frac{2}{c^2}\int_{\cal CC} d x^{i}
 (\vec{\omega}^{\prime} \times \vec{x})^i\nn
\\
&=-\frac{2}{c^2}\int d \vec{{\cal A}}\cdot 
\Big[(\vec{x}\cdot\nabla)\vec{\omega}'-(\vec{\omega}'\cdot\nabla) \vec{x}
+\vec{\omega}' (\nabla \cdot\vec{x})-\vec{x}(\nabla\cdot\vec{\omega}')\Big],\label{Totalframedragingeffect1}
\el 
where $\int_{\cal CC} d x^{i}
(\vec{\omega}^{\prime} \times \vec{x})^i
=-\int_{\cal C} d x^{i}
(\vec{\omega}^{\prime} \times \vec{x})^i$ 
and the Stokes' theorem have been used in the first and second 
equalities, respectively. Since $\vec{\omega}'$ is a constant vector in the 
laboratory, the first and second terms of \eqref{Totalframedragingeffect1} vanish, yielding
\bl
(\delta T_1-\delta T_2)_{FD}
&=-\frac{2}{c^2}\int d \vec{{\cal A}}\cdot 
\Big[-(\vec{\omega}'\cdot\nabla) \vec{x}
+\vec{\omega}' (\nabla \cdot\vec{x})\Big],\label{Totalframedragingeffect1afterchange}
\el 
where the spatial gradient is defined as $\nabla=\hat{e}_x^i\partial_i$. 
The angular velocity $\vec{\omega}'$ in \eqref{rotation} is contributed by 
$\vec{\omega}_{SA}^{\prime}$,
$\vec{\omega}'_{TP}$,
$\vec{\omega}'_{Geo}$ and 
$\vec{\omega}'_{LT}$. 
Therefore, the time delay due to the frame dragging effect
can be divided into four parts, which will be calculated below.

The angular
velocity of the Sagnac effect is 
\bl
\vec{\omega}_{SA}^{\prime}\equiv \vec{\omega}
\big(1+\frac{1}{2} \frac{v^2}{c^2}+\frac{U}{c^2}\big).\label{SAAomega}
\el
Replacing \eqref{SAAomega} 
with $\vec{\omega}'$ in 
\eqref{Totalframedragingeffect1afterchange}, one has 
\bl
(\delta T_1-\delta T_2)_{FD,SA}&=-\frac{4\vec{\cal A}\cdot \vec{\omega}}{c^2}
-\frac{2\vec{\cal A}\cdot \vec{\omega}}{c^4}
 \big((\omega R)^2\sin^2\theta-\frac{2G M}{R}\big),\label{1222}
\el
where \eqref{inverrelation3123}, \eqref{xintermodorighn}, and \eqref{intermsofvab} have been used. The second term of \eqref{1222} was previously overlooked in Refs. \cite{Brandy2021,Bosi2009}.

The angular velocity of the Thomas precession is 
\bl
\vec{\omega}'_{TP}\equiv\frac{1}{2}\frac{\vec{v}\times
\vec{\gamma}}{c^2}=
\frac{\omega}{2c^2}\Big[(R\omega\sin\theta)^2 (\cos\theta \hat{e}^1_x-\sin\theta\hat{e}^2_x)-\frac{G M}{R}\sin\theta\hat{e}^2_x\Big],\label{1222YP}
\el
where \eqref{intermsofv}, \eqref{intermsofacc}, and \eqref{intermsofgrav} have been employed.
Replacing $\vec{\omega}_{TP}^{\prime}$ with $\vec{\omega}^{\prime}$ in \eqref{Totalframedragingeffect1afterchange}, one obtains 
\bl
(\delta T_1-\delta T_2)_{FD,TP}
&=-\frac{4\vec{\cal A}\cdot\vec{\omega}'_{Tp}}{c^4},
\label{leadingoreerThom}
\el 
whose order is the same as 
that of the second term in \eqref{1222}.

The angular velocity of the Geodetic effect is 
\bl
\vec{\omega}'_{Geo}&\equiv\frac32(-\vec{v}\times \nabla U)=\frac3{2c^2}\Big[-\frac{G M\omega}{R}\sin\theta\hat{e}^2_x\Big],\label{138Geo}
\el
where \eqref{intermsofv} and \eqref{intermsofgrav} have been used.
Replacing $\vec{\omega}_{Geo}^{\prime}$ with $\vec{\omega}^{\prime}$ in \eqref{Totalframedragingeffect1afterchange}, one has 
\bl
(\delta T_1-\delta T_2)_{FD,Geo}&=\frac{4\vec{\mathcal{A}}\cdot\vec{\omega}'_{Geo}}{c^4}.
\label{leadingoreerGeo}
\el
The angular velocity of the 
Lense-Thirring effect is 
\bl
\vec{\omega}'_{LT}&\equiv\frac{GI\omega}{c^2R^3}\Big[ -2\cos\theta \hat{e}_{x}^1-\sin\theta \hat{e}_{x}^2\Big],\label{omegalt}
\el
where \eqref{intermsofavecab} is used. Replacing $\vec{\omega}_{LT}^{\prime}$ with $\vec{\omega}^{\prime}$ in \eqref{Totalframedragingeffect1afterchange}, one has 
\bl
(\delta T_1-\delta T_2)_{FD,LT}&=-\frac{4\vec{\mathcal{A}}\cdot\vec{\omega}'_{LT}}{c^4}.
\label{leadingoreerLT}
\el
Summing up \eqref{1222}, \eqref{leadingoreerThom},
\eqref{omegalt}, and \eqref{leadingoreerLT}
yields \eqref{aboveformula}.

\subsection{Paths ${\cal CC}: ABC$
and ${\cal C}: ADC$}\label{Appas}

Along the paths ${\cal CC}: ABC$ and ${\cal C}: ADC$, both the frame dragging and redshift effects contribute to the time delay. We shall calculate these effects below. To simplify the notation, we introduce
\begin{subequations}\bl
\vec{L}_{AB}=-\vec{L}_{CD}=L_{AB} f_{1,i}\hat{e}_{x}^i,\label{simdefinBA}
\\
\vec{L}_{BC}=-\vec{L}_{DA}=L_{BC} f_{2,i}\hat{e}_{x}^i,\label{simdefinBC}
\el\end{subequations}
where $L_{AB}$ and $L_{BC}$ denote the arm lengths of $AB$ and $CD$, and $BC$ and $DA$, respectively, and the unit vectors $\hat{L}_{AB}=f_{1,i}\hat{e}_{x}^i$ and $\hat{L}_{BC}=f_{2,i}\hat{e}_{x}^i$ are defined according to \eqref{uniCB} and \eqref{uniBA}. Therefore, the parameter functions for the arms $AB$, $BC$, $CD$, and $DA$ are 
\begin{subequations}\bl
x^i&= f_{1,i}p,\label{parameterAB}
\\
x^i&=f_{2,i}P+L_{AB}f_{1,i} ,\label{eqautionCB}
\\
x^i&=f_{1,i}p+L_{BC} f_{2,i},\label{eqautionDC}
\\
x^i&=f_{2,i}P, \label{eqautionDA}
\el\end{subequations}
where $0\leqslant p\leqslant L_{AB}$ and $0\leqslant P\leqslant L_{BC}$.

{\bf Frame dragging time delay}

The frame dragging time delay is given by the first term of \eqref{phasetotime11}
\bl 
 (\delta T_1-\delta T_2)_{FD}
 &=-\frac{1}{c^2}\int_{AB} d x^{i}
 (\vec{\omega}^{\prime} \times \vec{x})^i
 -\frac{1}{c^2}\int_{BC} d x^{i}
 (\vec{\omega}^{\prime} \times \vec{x})^i
 +\frac{1}{c^2}\int_{AD} d x^{i}
 (\vec{\omega}^{\prime} \times \vec{x})^i
 +\frac{1}{c^2}\int_{DC} d x^{i}
 (\vec{\omega}^{\prime} \times \vec{x})^i,\nn
 \\
 &=-\frac{1}{c^2}\int_{BC} d x^{i}
 (\vec{\omega}^{\prime} \times \vec{x})^i
 +\frac{1}{c^2}\int_{DC} d x^{i}
 (\vec{\omega}^{\prime} \times \vec{x})^i,\label{FDfinalequation2}
 \el
where the paths ${\cal CC}$ and ${\cal C}$ are divided into $AB$, $BC$ and $AD$, $DC$, respectively. The integrals along the paths $AB$ and $AD$ vanish in \eqref{FDfinalequation2} because $dx^i$ is proportional to $x^i$, where $x^i$ represents the coordinates along the arms $AB$ and $AD$, and $dx^i$ denotes the infinitesimal coordinate  intervals  along the light path. Replacing $\vec{\omega}'$ with $\vec{\omega}_{SA}$, $\vec{\omega}_{TP}$, $\vec{\omega}_{Geo}$, and $\vec{\omega}_{LT}$ in \eqref{FDfinalequation2}, respectively, one can then calculate the various time delays as follows.

For the Sagnac effect, one has 
\bl
-\frac{1}{c^2}\int_{BC} d x^{i}
 (\vec{\omega}_{SA}^{\prime} \times \vec{x})^i&=\frac{1}{c^2}\int_{DC} d x^{i}
 (\vec{\omega}_{SA}^{\prime} \times \vec{x})^i\nn
 \\
 &=\frac{{\cal A}\omega}{c^2}\big(1+\frac{1}{2}
  \frac{v^2}{c^2}+\frac{U}{c^2}\big)[ 
  (f_{1,2}f_{2,3}-f_{1,3}f_{2,2}) \cos \theta 
+(f_{1,1}f_{2,3}-f_{1,3} f_{2,1}) \sin\theta],\label{FDAG}
\el
where \eqref{SAAomega}, \eqref{simdefinBA}, \eqref{simdefinBC}, \eqref{eqautionCB} and  \eqref{unitarea}
have been used. Plugging \eqref{FDAG} into \eqref{FDfinalequation2}, one has 
\bl
(\delta T_1-\delta T_2)_{FD,SA}=-\frac{2\vec{{\cal A}}\cdot\vec{\omega'}_{SA}}{c^2},\label{adddsa}
\el
where \eqref{SAAomega}, \eqref{intermsofgeoab} and \eqref{unitarea} have been used.

For the Thomas precession effect, one has  
\bl
-\frac{1}{c^2}&\int_{BC} d x^{i}
 (\vec{\omega}_{TP}^{\prime} \times \vec{x})^i=\frac{1}{c^2}\int_{DC} d x^{i}
 (\vec{\omega}_{TP}^{\prime} \times \vec{x})^i\nn
 \\
 &=-\frac{\omega {\cal A}}{2c^2}
 \Big[(R\omega\sin\theta)^2 \big( (f_{2,1}f_{1,3}- f_{2,3}f_{1,1})\sin \theta 
 +(f_{2,2}f_{1,3}-f_{2,3} f_{1,2})\cos \theta\big)-\frac{G M}{R}\sin\theta 
 (f_{2,3} f_{1,1}-f_{2,1} f_{1,3} )\Big],\label{FDTP}
\el
where \eqref{1222YP}, \eqref{simdefinBA}, \eqref{simdefinBC}, \eqref{eqautionCB}
have been used. Plugging \eqref{FDTP} into \eqref{FDfinalequation2}, one has 
\bl
(\delta T_1-\delta T_2)_{FD,TP}=-\frac{2 \vec{{\cal A}}\cdot\vec{\omega}_{TP}'}{c^4},\label{adddtp}
\el
where \eqref{1222YP} and \eqref{intermsofgeoab} and \eqref{unitarea} have been used.

For the Geodetic effect, one has 
\bl
-\frac{1}{c^2}\int_{BC} d x^{i}
 (\vec{\omega}_{Geo}^{\prime} \times \vec{x})^i&=\frac{1}{c^2}\int_{DC} d x^{i}
 (\vec{\omega}_{Geo}^{\prime} \times \vec{x})^i=\frac{3\omega {\cal A}}{2c^2}
 \frac{G M}{R}\sin\theta 
 (f_{2,3} f_{1,1}-f_{2,1} f_{1,3} ),\label{FDGeo}
\el
where \eqref{138Geo}, \eqref{simdefinBA}, \eqref{simdefinBC} and \eqref{eqautionCB} have been used.
Plugging \eqref{FDGeo} into \eqref{FDfinalequation2}, one has 
\bl
(\delta T_1-\delta T_2)_{FD,Geo}=-\frac{2\vec{{\cal A}}\cdot\vec{\omega}_{Geo}'}{c^2},\label{addgeo}
\el
where \eqref{138Geo}, \eqref{intermsofgeoab} and \eqref{unitarea} have been used.

For the Lense-Thirring effect, one has 
\bl
-\frac{1}{c^2}\int_{BC} d x^{i}
 (\vec{\omega}_{LT}^{\prime} \times \vec{x})^i&=\frac{1}{c^2}\int_{DC} d x^{i}
 (\vec{\omega}_{LT}^{\prime} \times \vec{x})^i\nn
\\
& =-\frac{GI\omega{\cal A}}{c^2R^3}
\Big[  (f_{2,1}f_{1,3}- f_{2,3}f_{1,1})\sin \theta 
-2(f_{2,2}f_{1,3}-f_{2,3} f_{1,2})\cos \theta\Big],,\label{FDLT}
\el
where Eqs.\eqref{omegalt}, \eqref{simdefinBA}, \eqref{simdefinBC}, \eqref{eqautionCB}
have been used. Plugging \eqref{FDLT} into \eqref{FDfinalequation2}, one has 
\bl
(\delta T_1-\delta T_2)_{FD,LT}=-\frac{2\vec{{\cal A}}\cdot\vec{\omega}_{LT}'}{c^2},\label{addlt}
\el
where \eqref{omegalt}, \eqref{intermsofgeoab} and \eqref{unitarea} have been used.

{\bf Redshift time delay}

The redshift time delay corresponds to the second term of \eqref{0thapp21}. The infinitesimal interval $dl$ along the light path can be expressed as $(\hat{k} \cdot d\vec{x})$, where $\hat{k}$ is the unit tangent vector to the photon's path. Then the difference in the redshift time delay between the paths ${\cal CC}$ and ${\cal C}$ is
 \bl 
 (\delta T_1-\delta T_2)_{RS}&
 =\frac{1}{c^3}\int_{\cal CC} d\vec{x} \cdot\hat{k} ( \vec{\gamma}
 \cdot\vec{x})-\frac{1}{c^3}\int_{\cal C} d\vec{x} \cdot\hat{k} ( \vec{\gamma}
 \cdot\vec{x}),\nn
 \\
 &=\frac{1}{c^3}\int_{AB} d\vec{x} \cdot\hat{k} ( \vec{\gamma}
 \cdot\vec{x})
 +\frac{1}{c^3}\int_{BC} d\vec{x} \cdot\hat{k} ( \vec{\gamma}
 \cdot\vec{x})
 -\frac{1}{c^3}\int_{AD} d\vec{x} \cdot\hat{k} ( \vec{\gamma}
 \cdot\vec{x})
 -\frac{1}{c^3}\int_{DC} d\vec{x} \cdot\hat{k} ( \vec{\gamma}
 \cdot\vec{x})\label{finalequation2}
 \el
where the paths ${\cal CC}$ and ${\cal C}$ are divided into $AB$ and $BC$, and $AD$ and $DC$, respectively.
By substituting the parameter functions from \eqref{parameterAB}, \eqref{eqautionCB}, \eqref{eqautionDC}, and \eqref{eqautionDA} into the first, second, third, and fourth integration terms of \eqref{finalequation2}, we obtain
\begin{subequations}
\bl
(\delta T_1-\delta T_2)_{RS,AB}&=-\frac{L_{AB}^2}{c^3}\frac{GM}{R^2}f_{1,1}
-\frac{L_{AB}^2}{c^3} \omega^2 R\sin\theta (\sin\theta f_{1,1}+ \cos\theta f_{1,2}),\label{C13}
\\
(\delta T_1-\delta T_2)_{RS,BC}&=-\frac{L_{BC}^2 }{ c^3}\frac{ G M}{R^2}
\Big[f_{2,1}+2\frac{L_{AB}}{L_{BC}} f_{1,1} \Big]\nn
\\
&~~~~~~~~~~~~~~-
\frac{L_{BC}^2 }{ c^3} \omega ^2R\sin\theta\Big[(f_{2,1} \sin\theta 
 +f_{2,2}  \cos \theta )
 +2\frac{L_{AB}}{L_{BC}} (\sin \theta f_{1,1} 
 + \cos \theta f_{1,2} )\Big],\label{tiemeCBa}
 \\
 (\delta T_1-\delta T_2)_{RS,DC}&=
-\frac{L_{AB}^2}{c^3}\frac{GM}{R^2}[f_{1,1}+ 2\frac{L_{BC}}{L_{AB}} f_{2,1}],\nn
\\
&~~~~~~~~~~~~-\frac{L_{AB}^2}{c^3} \omega ^2R\sin \theta[ 
(f_{1,1} \sin \theta +f_{1,2} \cos \theta )
+2\frac{L_{BC}}{L_{AB}}  (f_{2,1} \sin \theta +f_{2,2} \cos \theta  )],\label{tiemeDCa}
\\
(\delta T_1-\delta T_2)_{RS,AD}&=-\frac{ L_{BC}^2 }{ c^3 }\frac{G M}{R^2}f_{2,1}-\frac{ L_{BC}^2 }{ c^3} \omega ^2R \sin\theta
(f_{2,1}\sin\theta +f_{2,2}\cos\theta),\label{tiemeDCam}
\el
\end{subequations}
where the subscripts $AB$, $BC$, $DC$, and $AD$ denote the redshift time delays associated with the corresponding arms.

\end{CJK}
\end{document}